\documentclass{aa}  

\usepackage{textcomp}
\usepackage{graphicx} %
\usepackage{amsmath} %
\usepackage{amssymb} %
\usepackage{bm} %
\usepackage{upgreek} %
\usepackage{IEEEtrantools} %
\usepackage{multirow}
\usepackage[colorlinks=true,allcolors=blue,urlcolor=blue]{hyperref}
\usepackage{txfonts}
\usepackage[normalem]{ulem} 
\usepackage{multirow}
\usepackage[dvipsnames]{xcolor}

\newcommand{\sect}[1]{\text{Sect.~\ref{#1}}}
\newcommand{\fig}[1]{\text{Fig.~\ref{#1}}}

\newcommand{\multitd}{\textsc{multi3d}}
\newcommand{\blob}{\textsc{balder}}

\newcommand{\stagger}{\textsc{stagger}}

\newcommand{\mtd}{\textlangle3D\textrangle}
\newcommand{\kms}{\mathrm{km\,s^{-1}}}

\newcommand{\lgeps}[1]{\log{\epsilon_{\mathrm{#1}}}}
\newcommand{\sh}{S_{\mathrm{H}}}

\newcommand{\lgr}{\log{\tau_{\mathrm{Ross}}}}
\newcommand{\dex}{\mathrm{dex}}
\newcommand{\triplet}{\ion{O}{I} {$777\,\mathrm{nm}$}~triplet}
\begin{document} 

\title{Inelastic O+H collisions and the 
\ion{O}{I} 777nm solar centre-to-limb variation}
\author{A.~M.~Amarsi\inst{1,2}
\and
P.~S.~Barklem\inst{3}
\and
M.~Asplund\inst{2}
\and
R.~Collet\inst{4}
\and
O.~Zatsarinny\inst{5}}
\institute{Max Planck Institute f\"ur Astronomy, K\"onigstuhl 17, 
D-69117 Heidelberg, Germany\\
\email{amarsi@mpia.de}
\and
Research School of Astronomy and Astrophysics, Australian National University,
Canberra, ACT 2611, Australia
\and
Theoretical Astrophysics, Department of Physics and Astronomy, 
Uppsala University, Box 516, SE-751 20 Uppsala, Sweden
\and
Stellar Astrophysics Centre, Department of Physics and Astronomy, Aarhus University, Ny Munkegade 120, DK-8000 Aarhus C, Denmark
\and
Department of Physics and Astronomy, Drake University, Des Moines, 
IA 50311, USA}

\abstract{The \triplet~is a key diagnostic
of oxygen abundances in the atmospheres of FGK-type stars;
however, it is sensitive to departures from 
local thermodynamic equilibrium (LTE).
The accuracy of non-LTE line formation calculations has
hitherto been limited by errors in the inelastic 
O+H collisional rate coefficients;
several recent studies have used the Drawin recipe,
albeit with a correction factor $\sh$~that is calibrated to
the solar centre-to-limb variation of the triplet.
We present a new model oxygen atom that incorporates
inelastic O+H collisional rate coefficients
using an asymptotic two-electron model 
based on linear combinations of atomic orbitals,
combined with a free electron model 
based on the impulse approximation.
Using a 3D hydrodynamic \stagger~model solar atmosphere
and 3D non-LTE line formation calculations, 
we demonstrate that this physically motivated approach
is able to reproduce the solar centre-to-limb variation
of the triplet~to $0.02\,\dex$, 
without any calibration of the inelastic
collisional rate coefficients or other free parameters.
We infer $\lgeps{O}=8.69\pm0.03$~from the triplet~alone,
strengthening the case for a low solar oxygen abundance.}


\maketitle
\section{Introduction}
\label{introduction}

Oxygen is the most abundant metal in the cosmos. It is
an important source of opacity and nuclear energy 
in stellar interiors
\citep[e.g.][]{2011ApJ...743...24S,2012ApJ...755...15V},
and its cosmic origins are well understood
\citep[e.g.][]{2012ceg..book.....M}.
Consequently, oxygen abundances are useful for understanding
and tracing the formation and evolution of
planets, stars, and galaxies 
\citep[e.g.][]{2016A&amp;A...590A..74B,
2016ApJ...831...20B,2016ApJ...827..126B,
2016AstL...42..734S,2016PASJ...68...32T,2016MNRAS.459.3282W}.
This makes it important to develop and  test tools 
with which to determine these abundances accurately.

The \triplet~is one of the most commonly
used diagnostics for oxygen abundances in FGK-type stars.
It is well established that 
local thermodynamic equilibrium (LTE) is a poor assumption 
for these lines
\citep[e.g.][]{1968SoPh....5..260A,1974A&amp;A....31...23S,
1979A&amp;A....71..178E,1991A&amp;A...245L...9K,
2000A&amp;A...359.1085P,2003A&amp;A...402..343T,
2009A&amp;A...500.1221F,2013AstL...39..126S};
photon losses mean that the predicted line strengths
are too weak when LTE is assumed.
Accurate abundance analyses 
require non-LTE line formation calculations 
that are based on three-dimensional (3D) hydrodynamic model atmospheres
\citep{1995A&amp;A...302..578K,2004A&amp;A...417..751A,
2009A&amp;A...508.1403P,2015MNRAS.454L..11A,
2015A&amp;A...583A..57S}.
Calculations on an extended grid of
3D hydrodynamic \stagger~model atmospheres
\citep{2013A&amp;A...557A..26M} have demonstrated that 
oxygen abundances determined from 1D LTE models
can be in error by as much as $0.7\,\dex$, with the largest errors
found in metal-rich turn-off stars
\citep{2016MNRAS.455.3735A}.

The main uncertainty in contemporary non-LTE models of 
the \triplet~lies in the treatment of
inelastic collisions of neutral oxygen with neutral hydrogen.
Usually non-LTE studies of \ion{O}{I}, and of most other species,
have employed the Drawin recipe.
This is based on the classical formula of \citet{thomson1912xlii} 
for ionisation by electron impact,
extended by \cite{1968ZPhy..211..404D,1969ZPhy..225..483D}~to 
the case of ionisation by neutral atom impact, 
extended further to excitation by 
\citet{1984A&amp;A...130..319S}, and later corrected by
\citet{1993PhST...47..186L}.

For \ion{Li}{I} \citep{2003PhRvA..68f2703B,2003A&amp;A...409L...1B},
\ion{Na}{I} \citep{2010PhRvA..81c2706B,2010A&amp;A...519A..20B},
and \ion{Mg}{I} \citep{2012PhRvA..85c2704B,2012A&amp;A...541A..80B},
comparisons with scattering  cross-sections based on
quantum chemistry calculations 
have revealed the Drawin rate coefficients
to be incorrect by several orders of magnitude.
The Drawin recipe, based on classical physics,
fails to describe the physical mechanism
as it is now understood for these three species:
 electron transfer at avoided ionic crossings
\citep[][]{2013PhRvA..88e2704B,2016A&amp;ARv..24....9B}.  
For this reason, the Drawin recipe does not capture the processes with the
highest rates, namely charge transfer and excitation involving spin-exchange
between nearby states.

A common approach to correcting for inadequacies in the Drawin recipe
is to scale the rate coefficients by an empirical factor $\sh$.
\citet{2004A&amp;A...423.1109A} suggested using the centre-to-limb variation
of the \triplet~to calibrate $\sh$.
The reasoning is that the sensitivity to the
inelastic hydrogen collisions, relative to
inelastic electron collisions, follows
the ratio $N_{\mathrm{H}}/N_{\mathrm{e^{-}}}$. This ratio increases
with height, thus inelastic hydrogen collisions are more influential
on the spectra emergent from the solar limb.
\citet{2009A&amp;A...508.1403P} used  high-quality data
of the centre-to-limb variation
to calibrate $\sh$, obtaining $\sh\approx0.85$~and a corresponding
low solar oxygen abundance of $\lgeps{O}\approx8.68$.
However, using the same spectra 
but a different calibration procedure for $\sh$, 
as well as a different model solar atmosphere
and line formation code, \citet{2015A&amp;A...583A..57S} independently 
obtained $\sh\approx1.7$~and a corresponding
higher solar oxygen abundance of $\lgeps{O}\approx8.77$.

In general, given a model atom with $N$~levels,
we cannot expect a single scaling factor $\sh$~to be able to correct all 
$\frac{1}{2}N\left(N-1\right)$~rate coefficients
because the errors in the Drawin recipe
can vary significantly from transition to transition;
in particular, the Drawin recipe makes no 
predictions for radiatively forbidden transitions.
As such, the $\sh$~parameter may be hiding 
other deficiencies in the models of
\citet{2009A&amp;A...508.1403P} and \citet{2015A&amp;A...583A..57S},
and the reliability of both of their results remains an open question.

Having a robust, physically motivated description of the 
inelastic O+H collisions that is able to reproduce
the solar centre-to-limb variation of the \triplet~would 
indicate a thorough understanding of the statistical equilibrium 
of \ion{O}{I}.  This would make the triplet~more trustworthy 
as an oxygen abundance diagnostic
in the Sun, and in other late-type stars.

Here, we propose a new description for the inelastic O+H collisions.
The description is based on the \citet{2016PhRvA..93d2705B}  asymptotic two-electron
model, which is based on linear combinations of atomic orbitals,
combined with the \citet{1991JPhB...24L.127K} free electron model,
which uses the impulse approximation.
While this description is indeed physically motivated, we caution that
the cross-sections calculated using 
the free electron model are rather uncertain,
especially for lower-lying states.
Nevertheless, we use 3D non-LTE line formation calculations 
using a large model atom and a new 3D hydrodynamic model solar atmosphere
to show that this description is able to reproduce
the solar centre-to-limb variation of the \triplet~very well.

The structure of this paper is as follows. We describe the 
3D non-LTE line formation calculations in \sect{method}.
We present the centre-to-limb variation
analysis of the \triplet~in \sect{analysis}.
We discuss the results in \sect{discussion}.
We summarise the main points in \sect{conclusion}.

\section{Model}
\label{method}

\subsection{Model solar atmosphere}
\label{methodatmosphere}

\subsubsection{Full 3D radiative-hydrodynamical model}
\label{methodatmospherefull}

\begin{figure*}
\begin{center}
\includegraphics[scale=0.31]{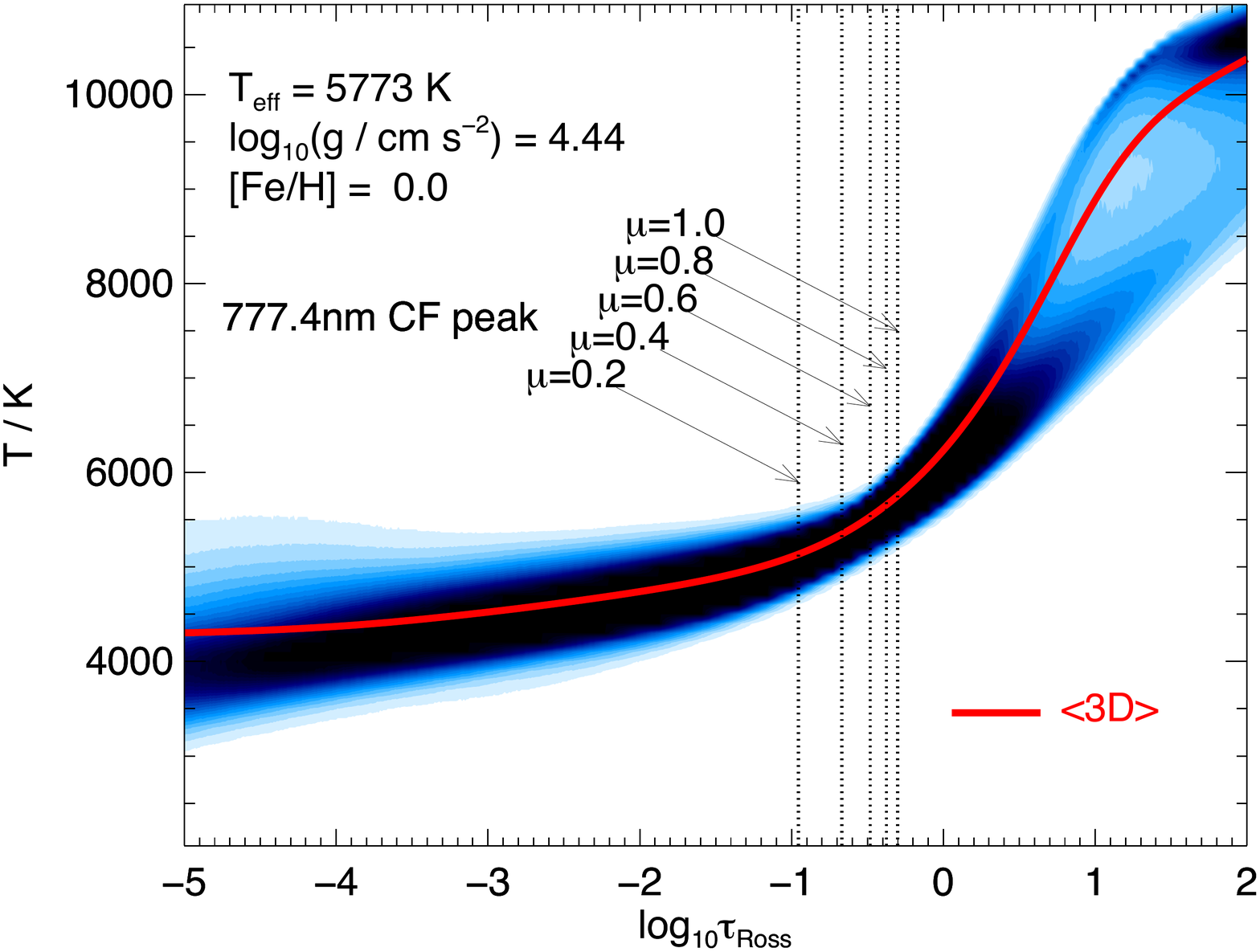}
\includegraphics[scale=0.31]{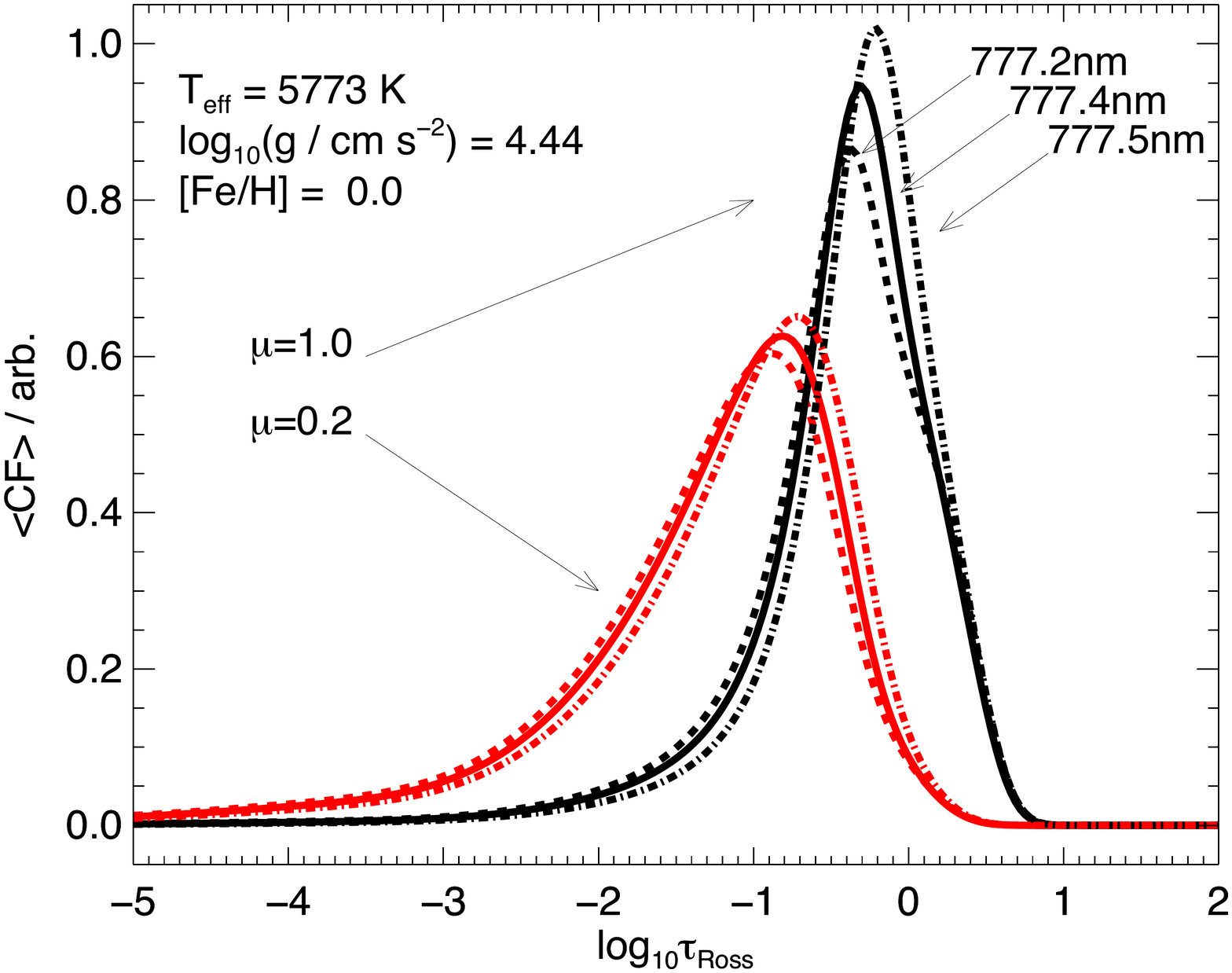}
\caption{Left: Temperature stratification of the
hydrodynamic model solar atmosphere used in this work;
also shown is the temperature stratification
of the temporally and horizontally averaged
\mtd~model solar atmosphere.
Vertical lines are used to
indicate the locations of the peaks
of the temporally  and horizontally averaged
contribution functions at different inclinations.
Right: Contribution functions to the absolute intensity depression
for the individual components of the \triplet~and at different
inclinations,
calculated on the full 3D model solar atmosphere and
subsequently temporally and horizontally averaged.
The contribution functions have been normalised individually,
such that the area under each is equal to unity.}
\label{fig:temp}
\end{center}
\end{figure*}

We illustrate the temperature stratification of
the 3D hydrodynamic model solar atmosphere used in this work
in \fig{fig:temp}. This is an updated version of the model used in
\citet{2009ARA&amp;A..47..481A} and \citet{2013A&amp;A...554A.118P},
and was recently used in \citet{2017MNRAS.468.4311L}
and \citet{2017A&amp;A...607A..75N}.
The model solar atmosphere was computed 
using a custom version of the \stagger~code 
\citep{Nordlund:1995,1998ApJ...499..914S}, tailored for the Sun.
The code solves the discretised equations for the conservation of 
mass, momentum, and energy coupled with the radiative transfer equation 
for a representative volume of the solar surface on a Cartesian mesh.

The numerical grid for the simulation covers $240^3$ grid points, 
spanning $6$~Mm in each horizontal direction and $4$~Mm vertically. 
The simulation domain completely 
covers the Rosseland mean optical depth
range ${-5}{\leq}\log_{10}\tau_{\mathrm{Ross}}{\leq}7$,
spanning about $8$ and $6$ pressure scale heights above 
and below the optical surface ($\tau_{\mathrm{Ross}}$=0), respectively.
Boundaries are open and transmitting 
at the top and bottom of the simulation, and
periodic in the horizontal directions. 
At the bottom boundary, located in the upper part of the solar 
convection zone, the entropy of the inflowing gas
and the gas pressure were set to constant values. 
A constant vertical gravitational acceleration with 
$\log_{10}\left(g / \mathrm{cm\,s^{-2}}\right)=4.44$~was enforced
throughout the simulation domain.

The simulation used an updated version of the 
equation of state by \citet{1988ApJ...331..815M} and of the continuous opacity 
package by \citet{1975A&amp;A....42..407G} 
(see \citealt{2013ApJ...769...18T}
and \citealt{2010A&amp;A...517A..49H} for a comprehensive 
list of the included continuous opacity sources). 
Sampled line opacities for wavelengths between $90\,\mathrm{nm}$ 
and $20000\,\mathrm{nm}$
came from B. Plez (priv. comm.) and \citet{2008A&amp;A...486..951G}.
The adopted chemical composition assumed for the calculation of the 
equation of state variables and opacities was taken from 
\citet{2009ARA&amp;A..47..481A}.

In order to compute the heating rates that enter the energy conservation
equation, the radiative transfer in the layers with
$\tau_{\mathrm{Ross}}{\leq}500$~were solved at each time step along rays 
crossing all grid points at the solar surface at nine 
inclinations (two $\mu$-angles and four $\phi$-angles, plus the vertical) 
using a Feautrier-like scheme \citep{1964CR....258.3189F}.
In the optically thick regions, the diffusion approximation was used
instead.

The computational load for the solution of the radiative transfer
was reduced by adopting the  opacity binning
(or multi-group opacity) method 
\citep{1982A&amp;A...107....1N,2000ApJ...536..465S}.
Prior to the calculations, wavelengths were sorted into different bins 
based on their spectral ranges  and on the strength 
of the associated opacity. 
Opacities in each bin were then appropriately averaged and the 
contributions from the monochromatic source functions were added together. 
During the simulation with the \stagger~code,
the radiative transfer equation was solved 
for the average opacities and cumulative source functions in 
each opacity bins. 
Coherent, isotropic continuum scattering was included 
in the source function in the optically thick
layers, and neglected in the optically thin layers.
The heating rates were then computed at all grid points 
by integrating the solution to the radiative transfer equation over the 
solid angle and over all opacity bins. 
We refer the reader to
Sect.~2.3 of \citet[][]{2011A&amp;A...528A..32C},
and in particular to the second approach presented there,
for further details on the radiative transfer scheme.

The effective temperature is not enforced; instead, we
fine-tuned the value of the specific entropy per unit mass of the 
inflowing gas at the bottom boundary to obtain a resulting effective
temperature value close to the observed one.
The convective simulation was run
for several convective turn-over timescales 
(about 21 hours of solar time)
to ensure that thermal and dynamical relaxation were achieved:
the mean effective temperature of the entire sequence is
$5773\,\mathrm{K}$, with standard deviation $16\,\mathrm{K}$.
This is very close to the reference solar value of $5772\,\mathrm{K}$
\citep{2016AJ....152...41P}.

\subsubsection{Averaged model}
\label{methodatmosphereaverage}

We also illustrate the temperature stratification of
an averaged 3D model solar atmosphere
(hereafter \mtd) in \fig{fig:temp}.
The sensitivity of the statistical equilibrium
on different aspects of the non-LTE modelling
was tested using this model solar atmosphere,
instead of the full 3D model solar atmosphere,
to save computational resources.
The \mtd~model was constructed by
calculating the mean of two thermodynamic quantities
in space (on surfaces of equal Rosseland mean optical depth)
and in time: the logarithmic gas density, and
the gas temperature to the fourth power.
All other quantities were then computed
consistently from these two quantities, the optical depth,
and the equation of state.

\subsubsection{Line-forming regions}
\label{methodatmospherelineformation}

It is interesting to briefly consider where  the \triplet~forms in the 
3D model solar atmosphere.
To that end, we illustrate in \fig{fig:temp}
contribution functions for the absolute intensity depression
at different inclinations, integrated over wavelength:
\phantomsection\begin{IEEEeqnarray}{rCl}
    \mathrm{CF}&\propto&
    \int{\alpha_{\lambda}\,S^{\text{eff}}_{\lambda}\,
    \mathrm{e}^{-\tau_{\lambda}}\,\mathrm{d}\lambda}\,,\\
    S^{\text{eff}}_{\lambda}&=&
    \left(\alpha^{\mathrm{l}}_{\lambda}/
    \alpha_{\lambda}\right)\left(I^{\mathrm{c}}_{\lambda}-
    S^{\mathrm{l}}_{\lambda}\right)\,
\end{IEEEeqnarray}
\citep[][integrand of Eq.~12]{2015MNRAS.452.1612A}.
The function was computed at every grid-point of the seven snapshots
(\sect{methodcode}) of the model solar atmosphere,
and subsequently temporally  and horizontally averaged 
on surfaces of equal Rosseland mean optical depth,
and area normalised.
These plots were calculated using 3D non-LTE radiative
transfer (\sect{methodcode}),
and using the model atom with
inelastic collisions with neutral hydrogen 
based on the asymptotic two-electron model of 
\citet{2017arXiv171201166B}
combined with the full, free electron model of
\cite{1985JPhB...18L.167K,kaulakys1986free,1991JPhB...24L.127K}
with an oxygen abundance of $8.7\,\dex$
(\sect{methodcollch}).

The \triplet~forms deep in the photosphere,
as expected from their
high excitation potentials.  
They form over an extended region; the contribution
functions are skewed to higher optical depths by the deep line
cores. 
For disk-centre profiles the peak in the contribution function is around
$-0.4\lesssim\lgr\lesssim-0.1$, depending on the line component
(the weaker components forming at greater optical depth),
with full widths at half maxima of around $\Delta\lgr\approx1.0$.
For profiles observed closer to the limb the contribution functions
shift to more optically thin regions, and also become broader. For example, at $\mu=0.2$~the peaks move to around 
$-0.9\lesssim\lgr\lesssim-0.6$,
with full widths at half maximum of around $\Delta\lgr\approx1.4$.
This broadening reflects the inhomogeneous nature
of the stable photosphere, with a  larger contribution
to the line formation
occurring in localised regions of higher gas temperature.

\subsubsection{Validation}
\label{methodatmospherevalidation}

\begin{figure*}
\begin{center}
\includegraphics[scale=0.31]{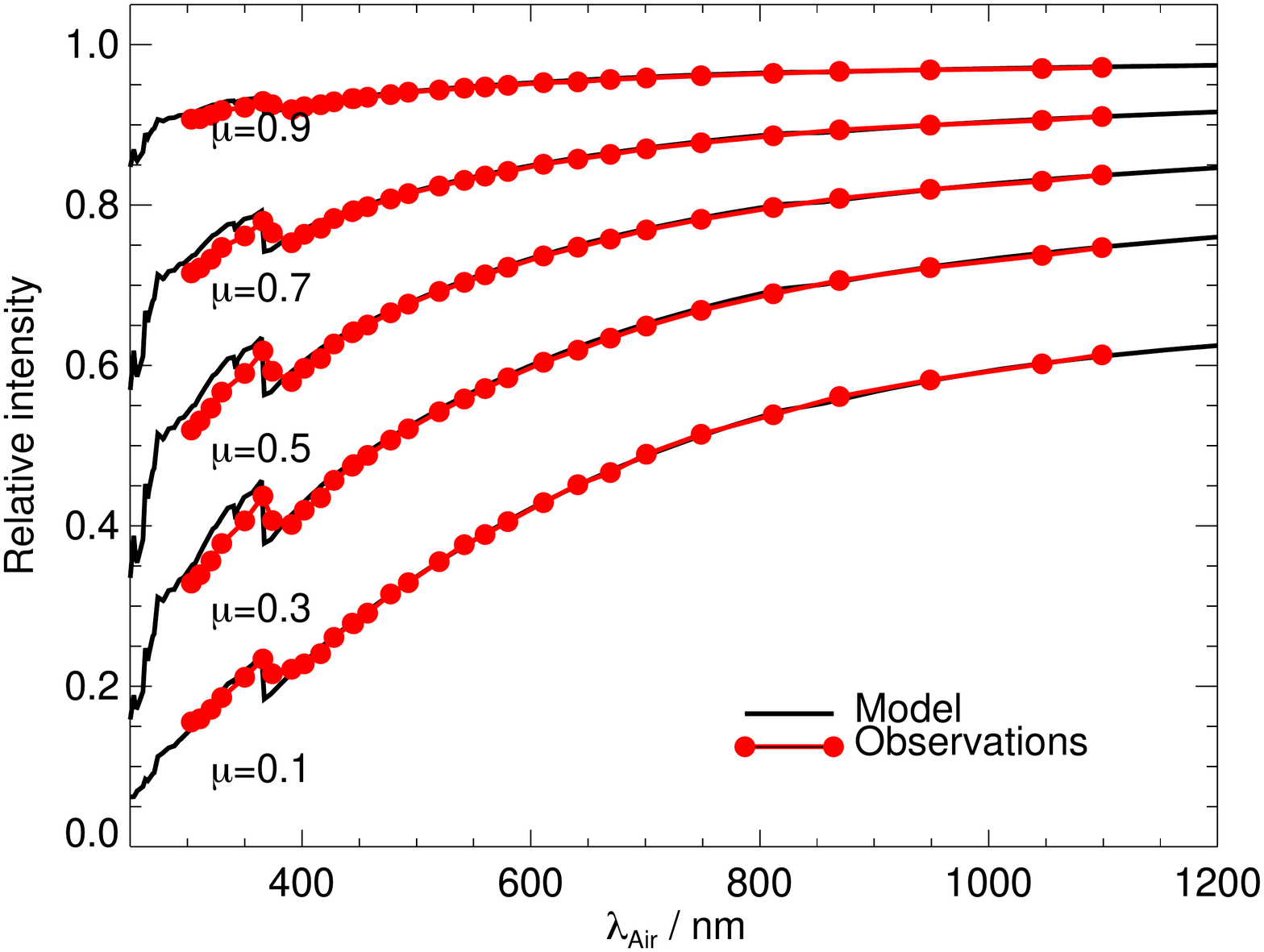}
\includegraphics[scale=0.31]{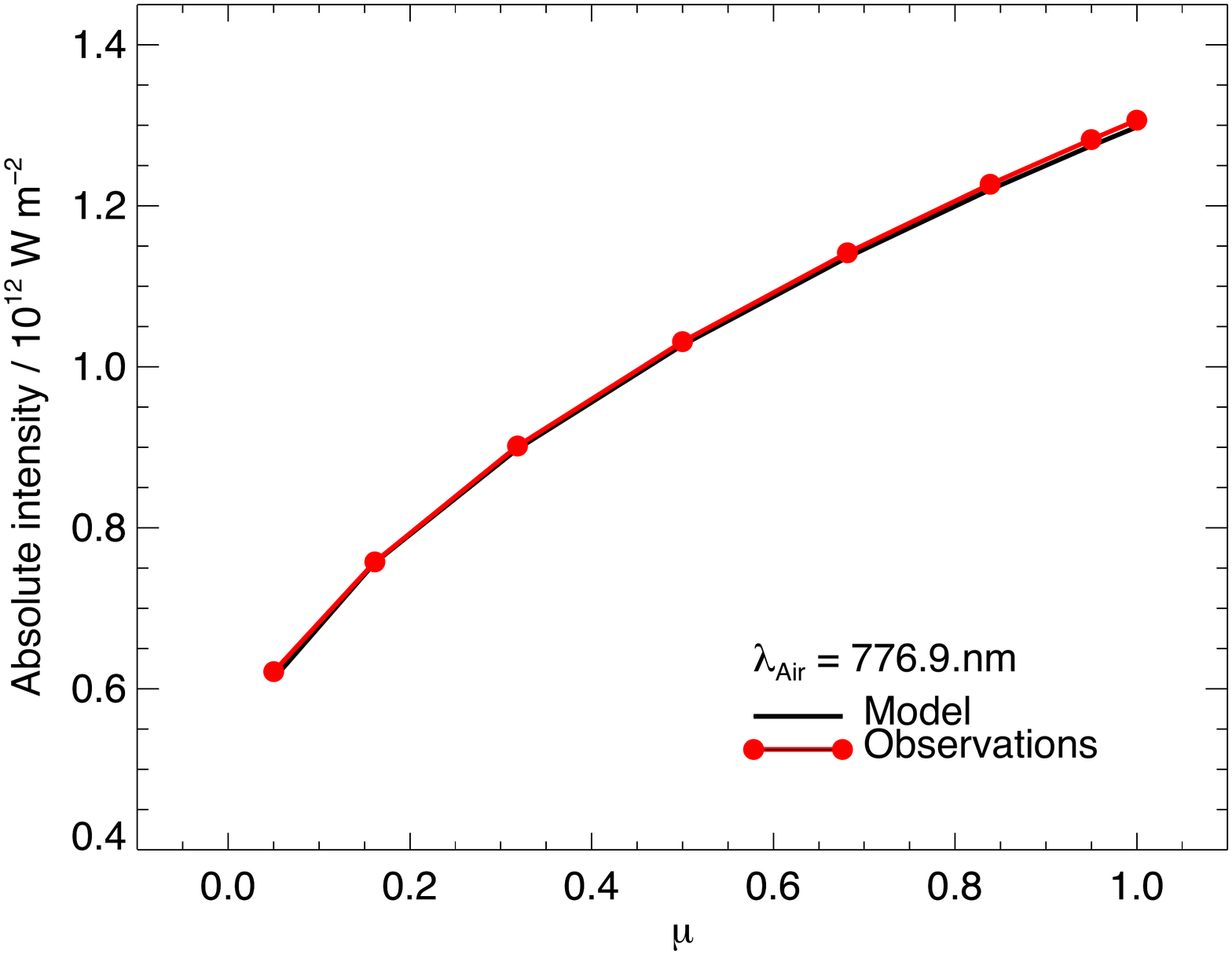}
\caption{Left: Relative continuum intensities
(normalised by the continuum intensity observed at disk-centre)
at various inclinations
as functions of wavelength, 
as predicted by the 3D hydrodynamic model solar atmosphere,
and as given by the observations of 
\citet{1994SoPh..153...91N}.
Right: Absolute continuum intensity in the vicinity of the \triplet~as
a function of inclination,
as predicted by the 3D hydrodynamic model solar atmosphere,
and as given by the observations of 
\citet{1994SoPh..153...91N},
put onto an absolute scale using the 
atlas of \citet{1984SoPh...90..205N}.}
\label{fig:cont_clv}
\end{center}
\end{figure*}

Before using the centre-to-limb variation
analysis to comment on the accuracy of the non-LTE modelling,
it is necessary to first verify that the model solar atmosphere is
an accurate representation of the quiet Sun.
    
In \fig{fig:cont_clv}~we illustrate 
the centre-to-limb
variation of the relative continuum intensity at different optical 
and near-infrared wavelengths,
as predicted by the 3D hydrodynamic model solar atmosphere,
and as given by the observations of \citet{1994SoPh..153...91N}.
We also compare the centre-to-limb
variation of the absolute continuum intensity (emergent
from the model solar atmosphere)
in the vicinity of the \triplet, 
putting the data of \citet{1994SoPh..153...91N} onto an absolute scale
using the disk-centre intensity atlas of \cite{1984SoPh...90..205N}
and a small continuum window centred on 
$\lambda_{\text{Air}}=776.885\,\mathrm{nm}$.

\fig{fig:cont_clv}~shows that
redward of the Balmer discontinuity, 
the observed centre-to-limb variation is very well reproduced
by the model, from disk-centre down to $\mu\approx0.1$. 
This is strong evidence that the 
temperature stratification is realistic in the 
continuum-forming regions
$-1.0\lesssim\log\tau_{\lambda}\lesssim0.0$.
Slight discrepancies at shorter wavelengths 
may indicate that some UV opacity is missing in our synthesis.
In addition, \fig{fig:cont_clv} shows that
the absolute continuum intensities are in good agreement, which mainly
verifies that the effective temperature of the model 
solar atmosphere is correct.

For further validation, we note that 
the 3D hydrodynamic model solar atmosphere used here  
is the same one used by \citet{2017MNRAS.468.4311L} to analyse
the centre-to-limb variations of ten \ion{Fe}{I} lines and one
\ion{Fe}{II}~line.  They were generally able to reproduce the 
the centre-to-limb variations down to $\mu\approx0.4$; 
disagreements closer to the limb 
can be explained by uncertainties in their non-LTE modelling.
These eleven lines have different
excitation potentials, oscillator strengths, and wavelengths;
and consequently, different regions of line formation.
We calculated the contribution functions for these lines
in the same way as we did for the
\triplet~(\sect{methodatmospherelineformation}), 
but under the assumption of LTE;
this is valid because the non-LTE effects are small for 
\ion{Fe}{I}~lines in the Sun.
For disk-centre profiles the peaks in the contribution function 
are in the range $-0.4\lesssim\lgr\lesssim-1.4$,
while for profiles observed at $\mu=0.4$~the peaks
are in the range $-0.9\lesssim\lgr\lesssim-1.6$.
This covers and extends beyond the region in which
the \triplet~forms.

Finally, we comment on magnetic fields,
which were neglected in our purely hydrodynamic simulations.
The \triplet~is sensitive to 
magnetic fields, both directly 
(via Zeeman broadening) and indirectly 
(via changes to the atmospheric structure),
as shown by the MHD simulations of
\citet{2012A&amp;A...548A..35F},
\citet{2015ApJ...802...96F}, 
\citet{2015ApJ...799..150M},
and \citet{2016A&amp;A...586A.145S}. 
The first two studies imposed a vertical mean field of 
$100\,\mathrm{G}$~in their MHD simulations.
This magnetic field topology has a large impact 
on the atmospheric structure, that promotes a strong 
indirect effect of magnetic fields on spectral line formation;
as such, their results suggest abundance corrections
can reach approximately $0.1\,\dex$,
compared to the purely hydrodynamic case. 
In contrast, the last two studies
adopted small-scale, self-dynamo models 
with mean strengths of $80$--$160\,\mathrm{G}$~in their MHD simulations.
Their results suggest much smaller abundance corrections;
for the \triplet~these are only of the order of $0.01\,\dex$.
Observations suggest that the  
small-scale dynamo approach is more realistic
\citep[e.g.][]{2004Natur.430..326T}.


\subsection{Line formation code}
\label{methodcode}

Our 3D non-LTE radiative transfer code 
\blob, which is a custom version of
\multitd~\citep{1999ASSL..240..379B,2009ASPC..415...87L},
was used in this study. We refer the reader to
\citet{2018arXiv180402305A}~and 
references therein for an overview of the code.

Calculations were performed across seven snapshots of the
3D hydrodynamic model solar atmosphere that we
described in \sect{methodatmospherefull}.
These snapshots were equally spaced in solar time.
Prior to performing line formation calculations,
the snapshots were resampled 
onto a mesh with $80\times80\times220$ grid points, as described in 
\citet{2018arXiv180402305A}. 
Line formation calculations were performed for 
each model snapshot for different values of oxygen abundance $\lgeps{O}$,
in steps of $0.2\,\dex$,
under the assumption that these variations have no effect on
the background atmosphere (i.e.~that oxygen is a trace element).

The calculations on the \mtd~model solar atmosphere
proceeded in mostly the same way as those
on the 3D hydrodynamic model solar atmospheres.
In general, 
microturbulent and macroturbulent broadening
\citep[e.g.][Chapter 17]{2008oasp.book.....G} need to be included
in analyses based on 1D model atmospheres
in order to reproduce the line broadening effects of 
the convective velocity field oscillations and temperature inhomogeneities
\citep{2000A&amp;A...359..729A}.
In this work a depth-independent microturbulence of 
$\xi=1\,\kms$~was adopted,
and macroturbulence was included by convolving the profiles
with Gaussian kernels of freely varying width
$\varv_{\mathrm{mac}}\left(\mu\right)$~so as to best fit the observed profile.

\subsection{Model atom}
\label{methodatom}

\begin{figure*}
\begin{center}
\includegraphics[scale=0.31]{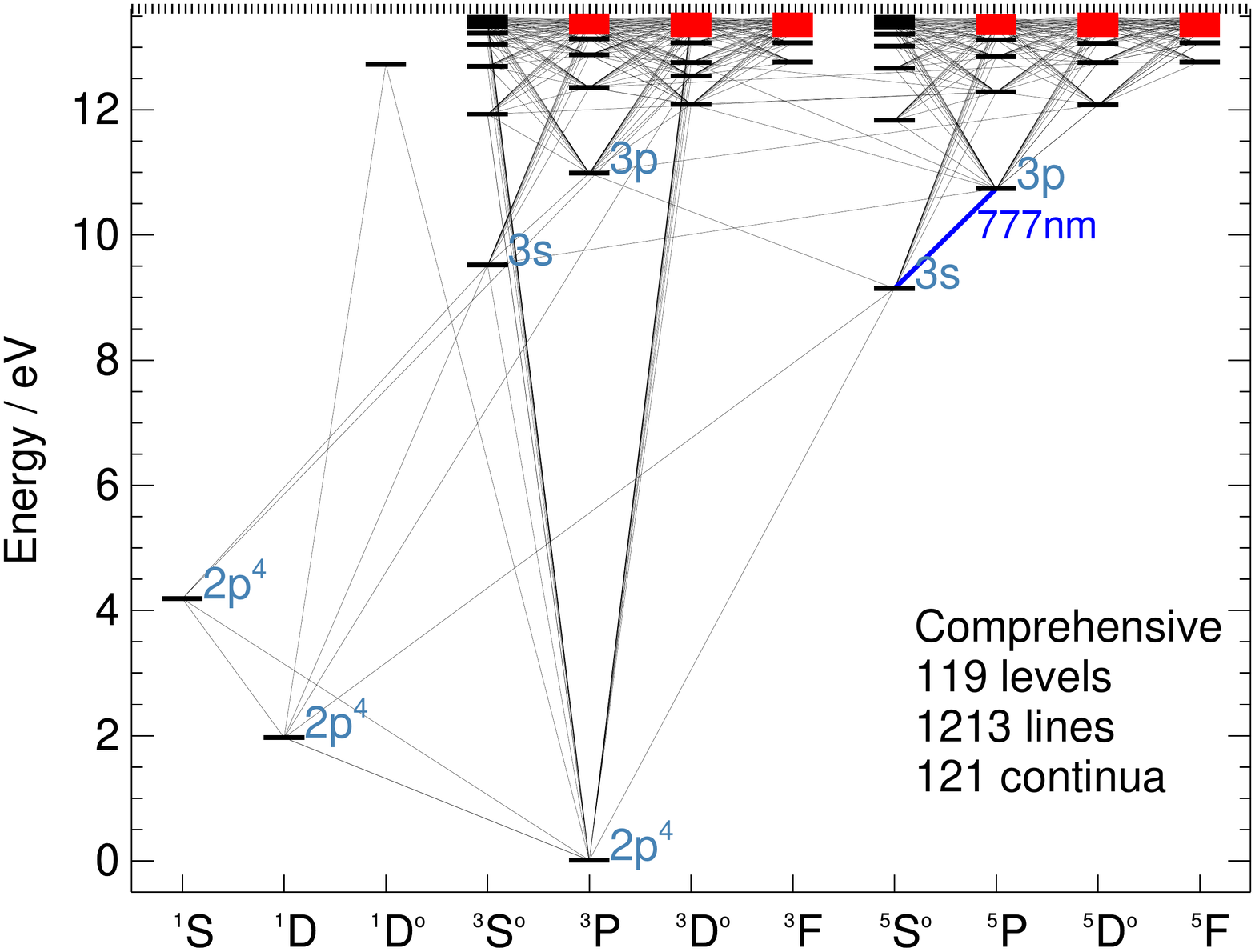}
\includegraphics[scale=0.31]{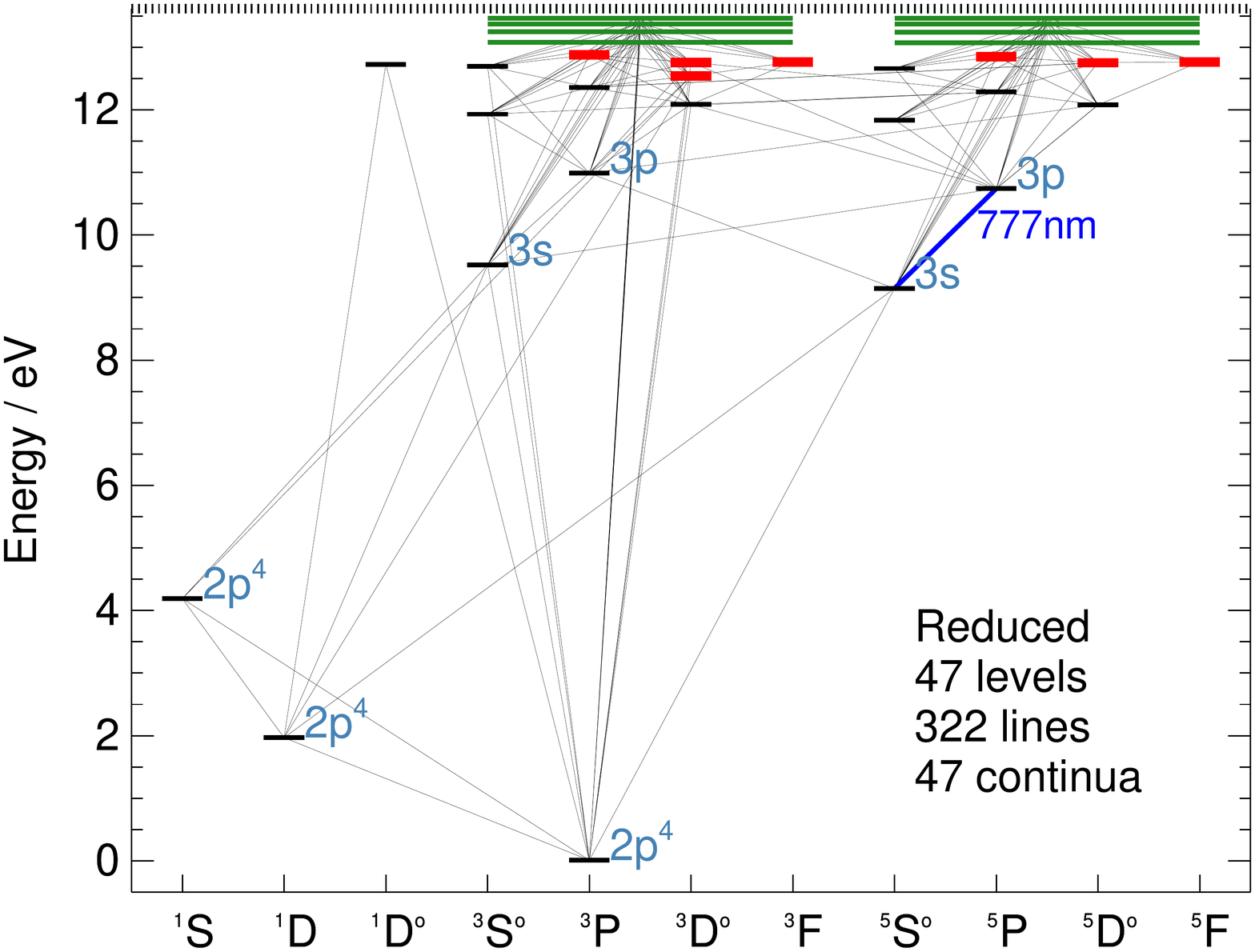}
\caption{Grotrian diagram of \ion{O}{I} in the 
comprehensive and reduced model oxygen atoms.
Super levels are depicted
as solid horizontal green lines.
Levels that do not resolve fine structure 
are depicted as thick red lines;
in the comprehensive atom, these correspond to 
levels taken from the Opacity Project.
The final reduced model atom
includes three  levels of \ion{O}{II} (fine structure collapsed),
bringing the total number of levels to 50.}
\label{fig:grotrian}
\end{center}
\end{figure*}

\begin{figure}
\begin{center}
\includegraphics[scale=0.31]{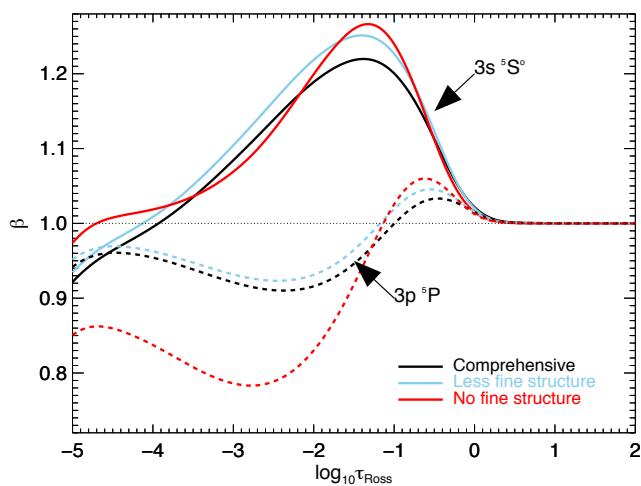}
\caption{Departure coefficients
for the lower and upper levels
of the \triplet~in the \mtd~model solar atmosphere.
The departure coefficients as predicted by the 
reduced model atom overlap with those predicted
by the comprehensive model atom in this plot.
Also shown are the departure coefficients
predicted from the comprehensive model atom but with all 
fine structure above the triplet~(i.e.~above $10.74\,\mathrm{eV}$)
collapsed (`Less fine structure'),
and  from the comprehensive model atom but with all 
fine structure down to and including
the ground level collapsed (`No fine structure'), 
illustrating the importance of resolving fine structure
in non-LTE model atoms.}
\label{fig:reduction}
\end{center}
\end{figure}

We illustrate two model atoms in \fig{fig:grotrian}:
a comprehensive model, which is more complete
than those used in previous studies
\citep[e.g.][]{2009A&amp;A...508.1403P,
2015A&amp;A...583A..57S,2016MNRAS.455.3735A},
and a reduced model, which was used for the actual calculations
at reduced computational cost but without significant loss of accuracy.

Experimental fine structure energies for 
\ion{O}{I} up to $13.18\,\mathrm{eV}$~above 
the ground state (up to and including $\mathrm{6p\,^{3}P}$),
and fine structure energies for \ion{O}{II} up to $18.64\,\mathrm{eV}$~above 
the ground state (up to and including $\mathrm{2p^{3}\,^{2}P^{o}}$),
were taken from the compilation of \citet{gallagher1993tables}
via the NIST Atomic Spectra Database \citep{NIST_ASD}.
These were supplemented with all available \ion{O}{I} energies
given in the Opacity Project online database
\citep[TOPbase;][]{1993A&amp;A...275L...5C}.

Experimental \ion{O}{I} oscillator strengths were taken from
\citet{1991JPhB...24.3943H}
via the NIST Atomic Spectra Database. 
These were supplemented with all available \ion{O}{I}
oscillator strengths in TOPbase.
The TOPbase data set 
does not include fine structure;
we therefore carefully split the oscillator strengths
under the assumption of pure LS coupling, using the tables given in 
\citet[][Sect.~27]{1973asqu.book.....A}.
Natural broadening coefficients
were calculated using the lifetimes given in TOPbase,
and van der Waals broadening coefficients, when possible,
were based on the theory of Anstee, Barklem, and O'Mara 
\citep[ABO;][]{1995MNRAS.276..859A,1997MNRAS.290..102B,
1998MNRAS.296.1057B}. 
Monochromatic photoionisation cross-sections were 
extracted from TOPbase.

The reduction broadly follows the procedure
described in \citet[][see Sect.~2.3.3 for details]{2017MNRAS.464..264A}.
First, eight \ion{O}{I} super levels were constructed,
starting at around $13.0\,\mathrm{eV}$~($\mathrm{6s\,^{5}S^{o}}$).
These super levels combine composite levels that share 
the same spin quantum number and are 
separated in energy by less than $0.1\,\mathrm{eV}$.
They were were constructed by
weighting the composite level energies according to their Boltzmann factors
(adopting a characteristic temperature of $5000\,\mathrm{K}$).
Affected lines and continua were collapsed into 
super lines and super continua.

Second, we considered collapsing the fine structure in the model atom.
All fine structure in the \ion{O}{II} system was collapsed;
however, some care has to be taken with fine structure in \ion{O}{I}
\citep[e.g.][Appendix B]{2015A&amp;A...583A..57S}.
Collapsing fine structure in \ion{O}{I}
can significantly impact the predicted departure coefficients;
we illustrate this, using the \mtd~model 
solar atmosphere, in \fig{fig:reduction}.
Although these differences
ultimately only have  a small impact on the 
\triplet~in 1D model atmospheres
\citep[e.g.][]{1993A&amp;A...275..269K},
the impact may be larger on the line profiles
emergent from 3D hydrodynamic model atmospheres.
We therefore chose to be cautious, and 
resolved fine structure in \ion{O}{I} levels
up to around $12.5\,\mathrm{eV}$~(up to and including $\mathrm{4p\,^{3}P}$).

\subsection{Inelastic collisions}
\label{methodcoll}

\begin{figure*}
\begin{center}
\includegraphics[scale=0.31]{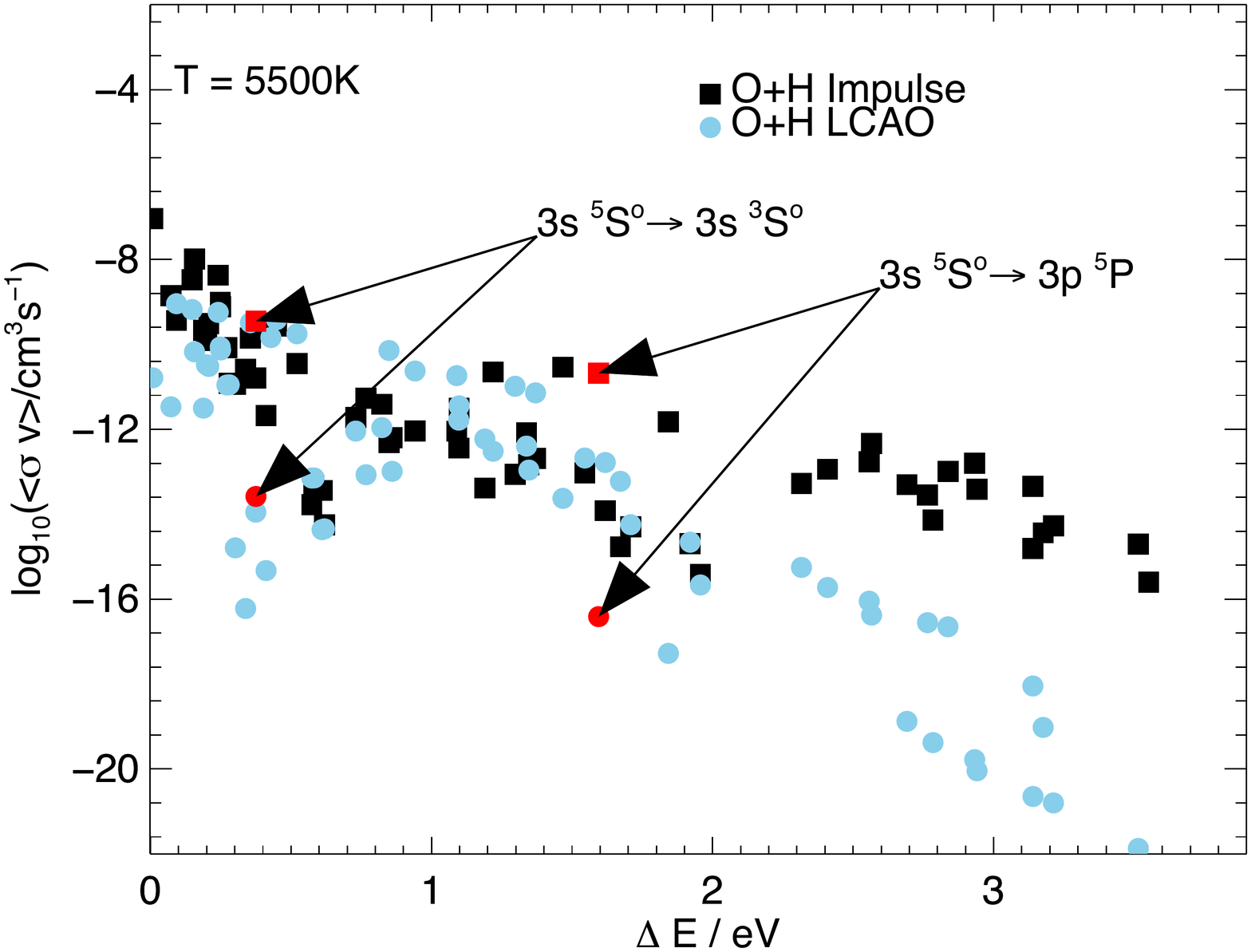}\includegraphics[scale=0.31]{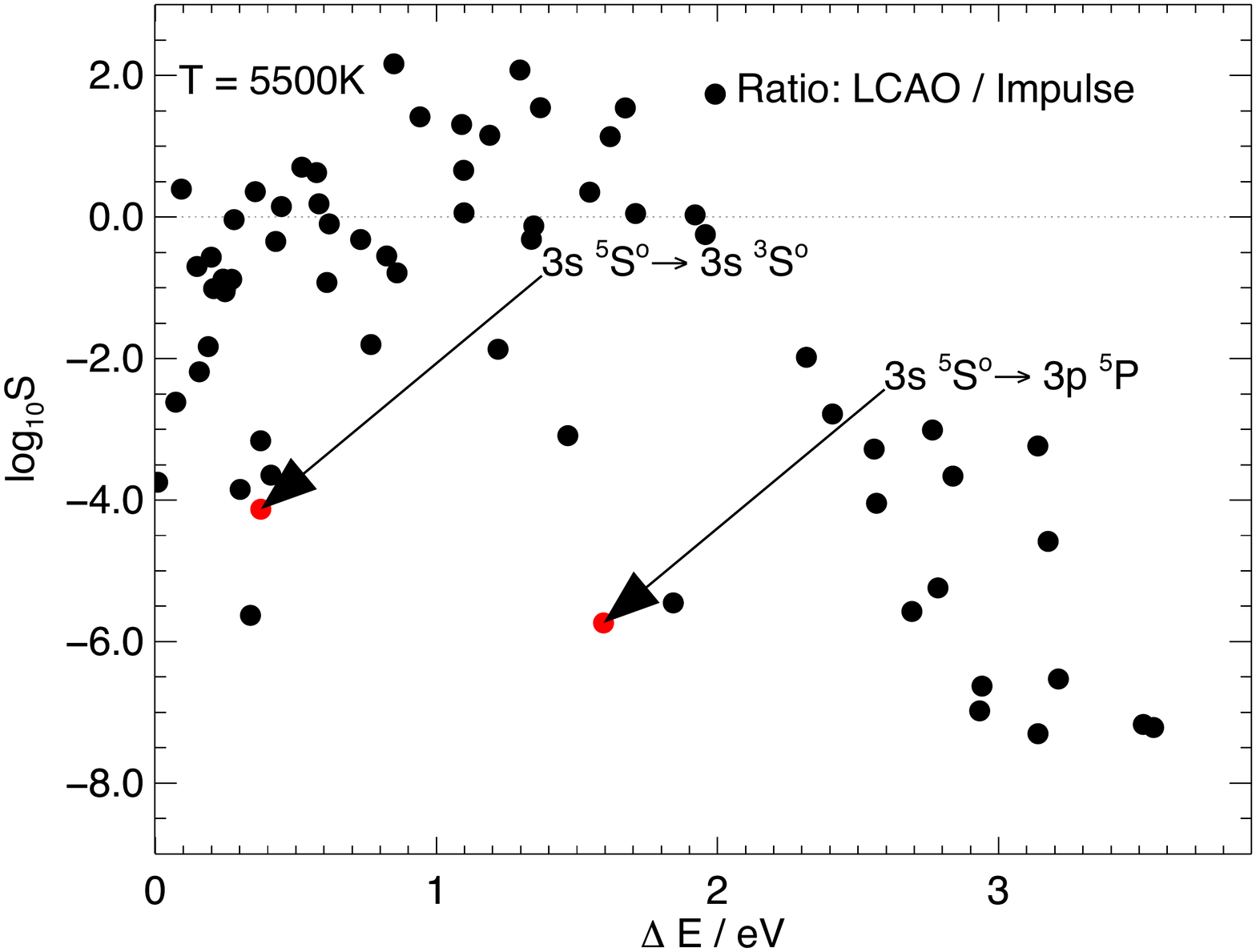}
\includegraphics[scale=0.31]{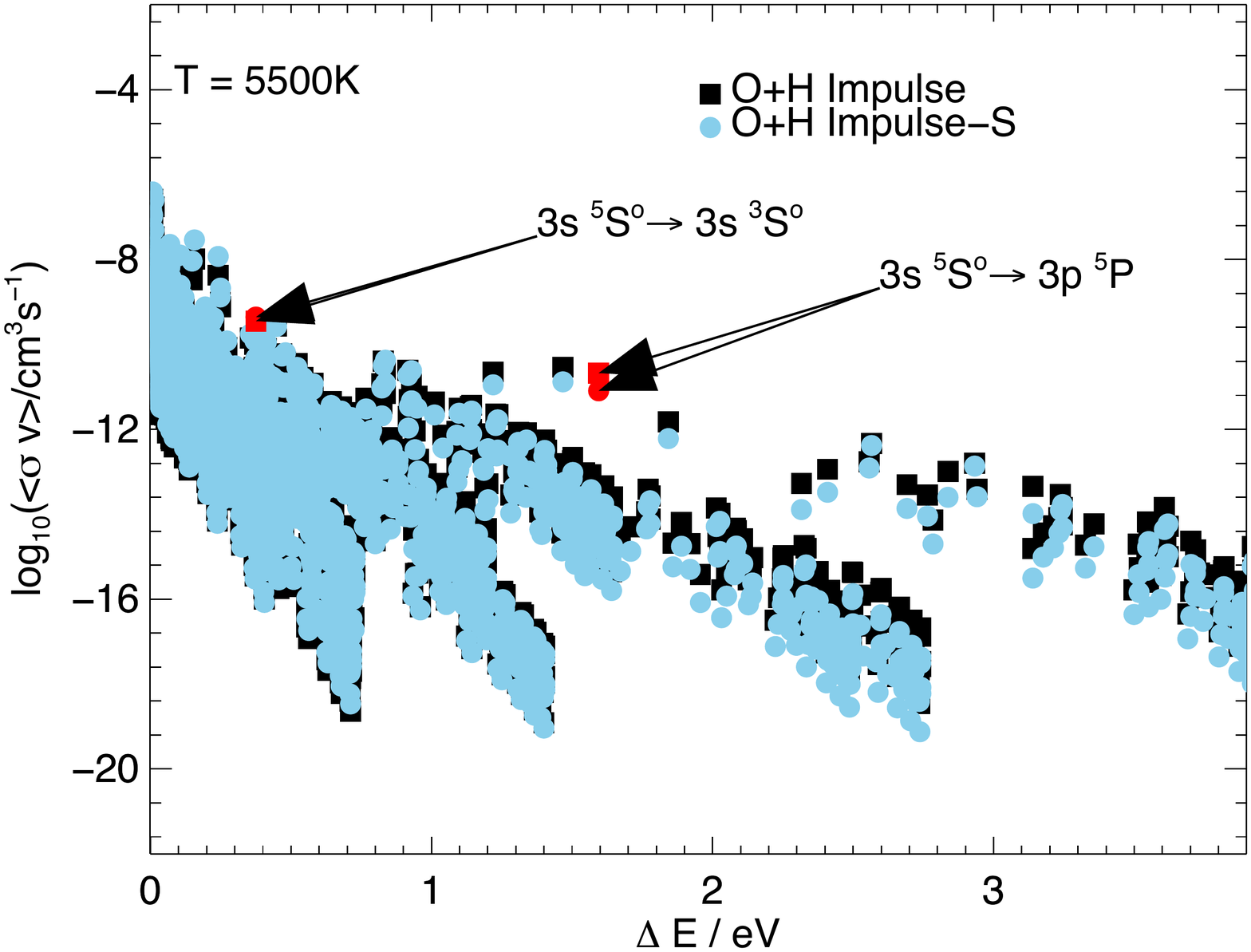}\includegraphics[scale=0.31]{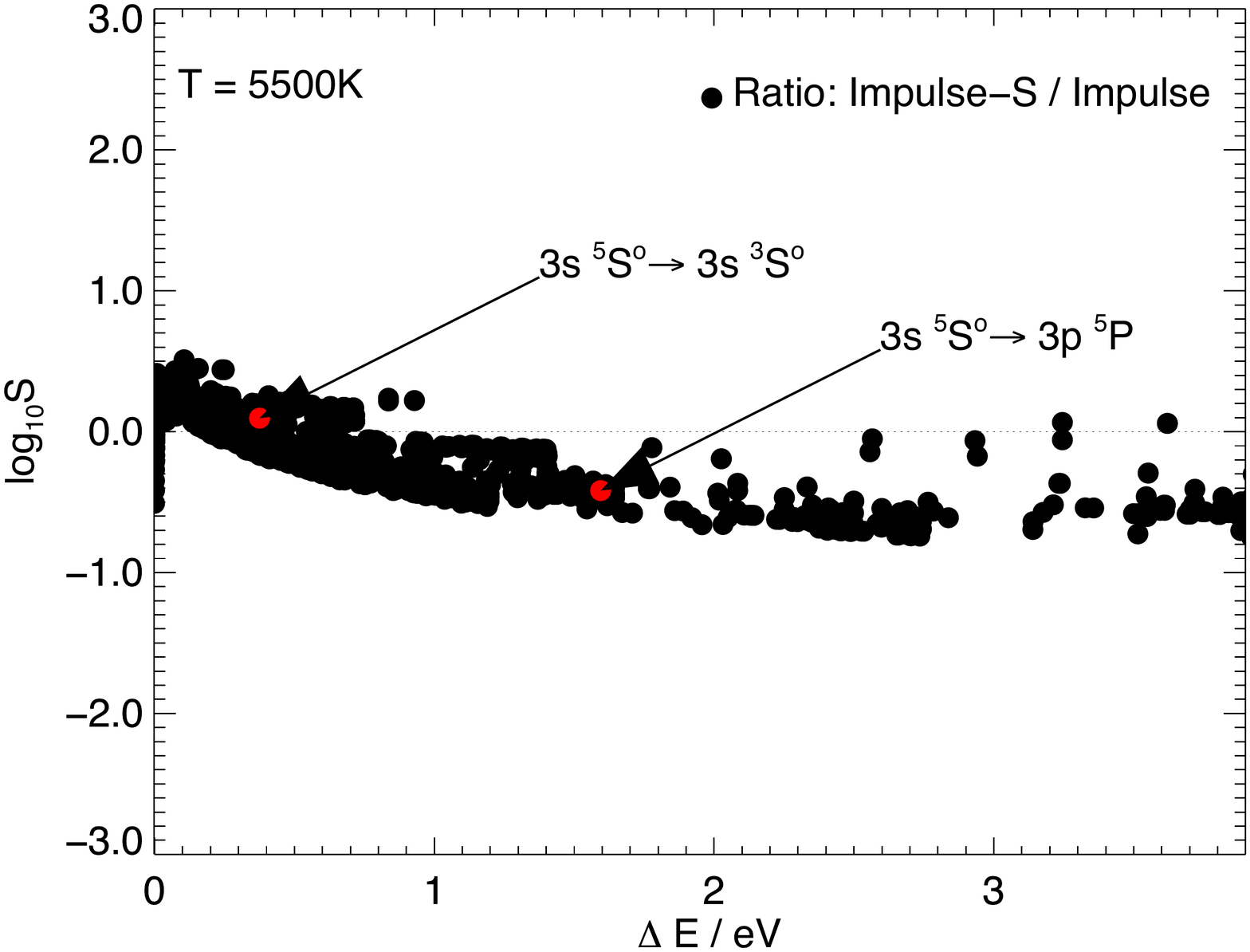}
\includegraphics[scale=0.31]{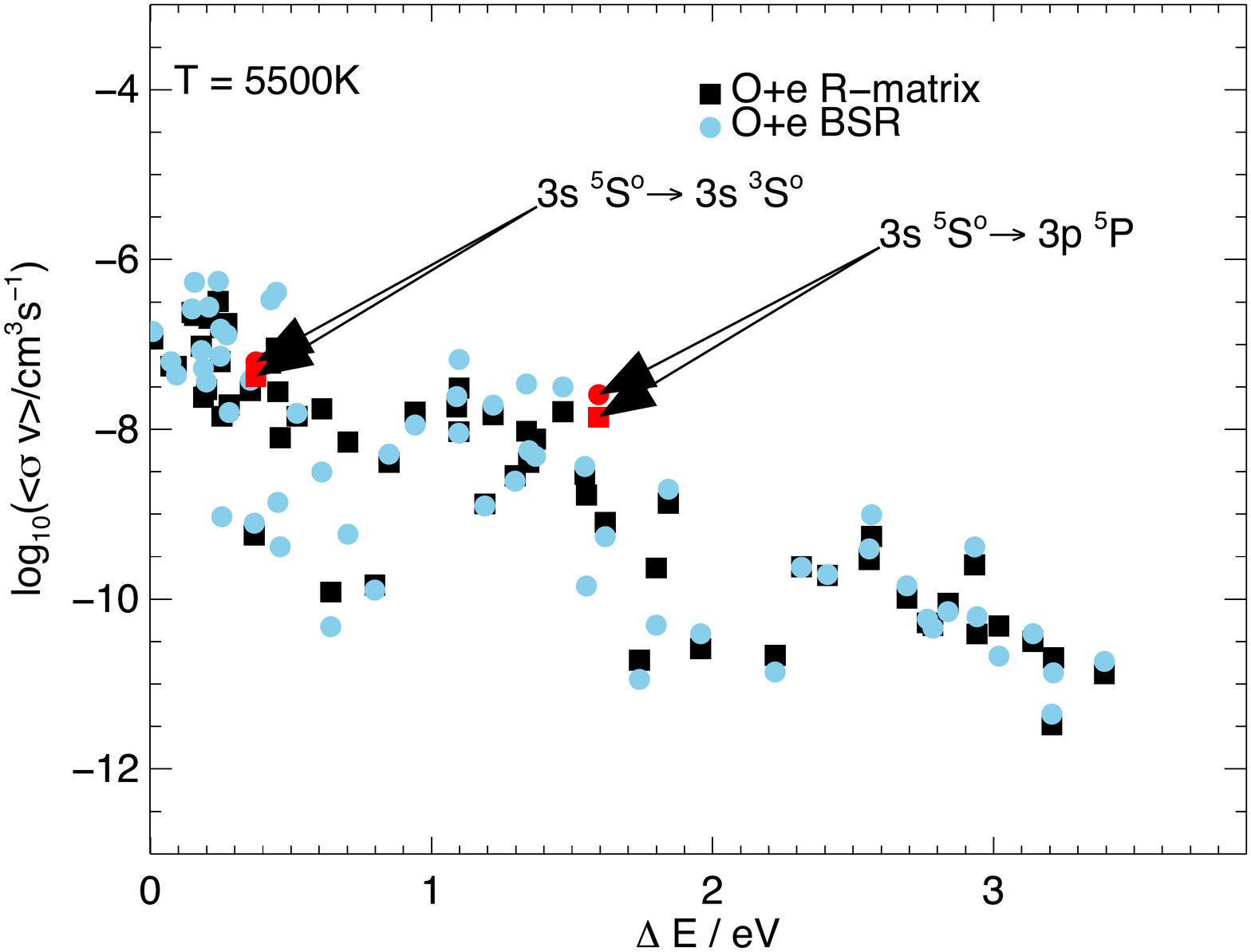}\includegraphics[scale=0.31]{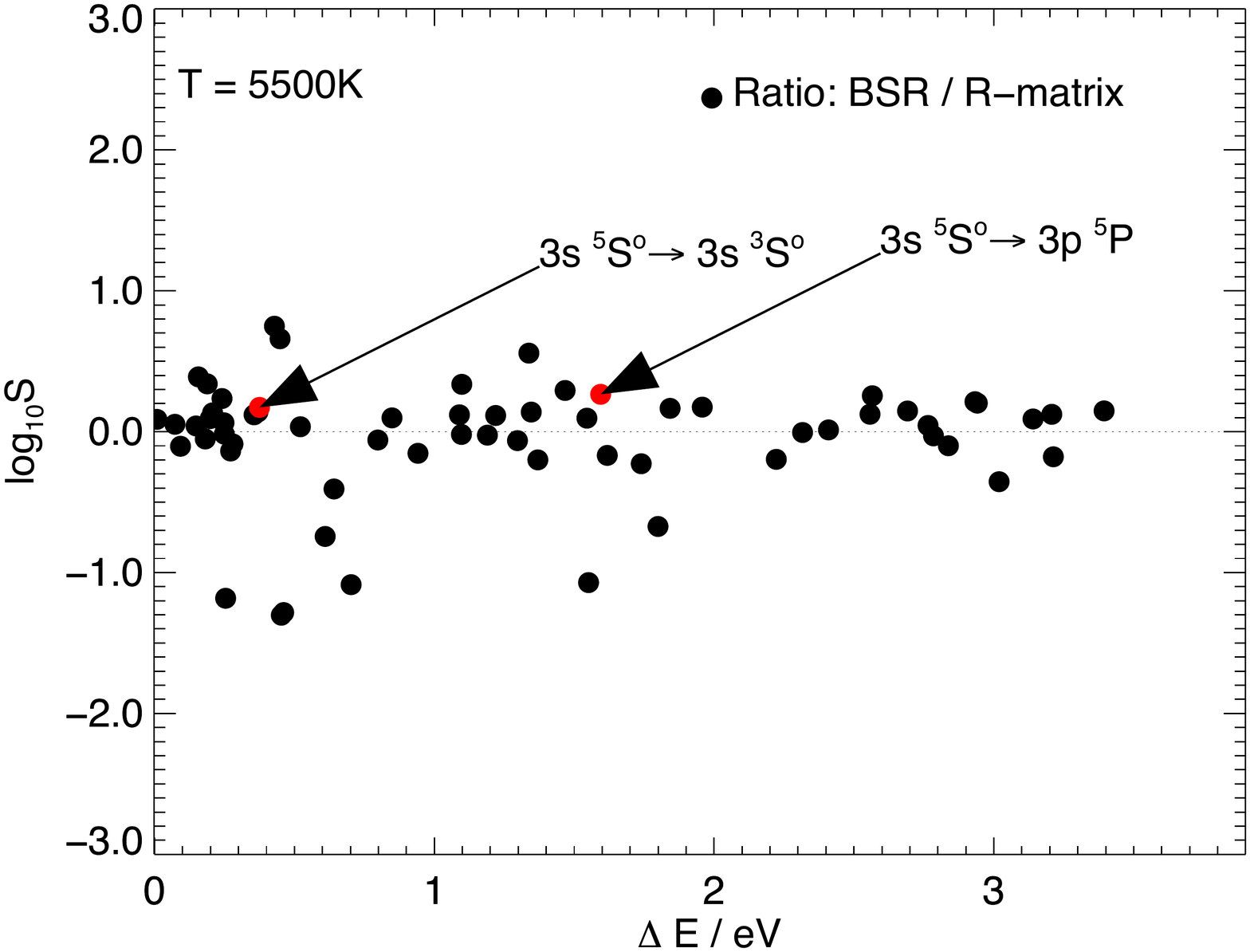}
\caption{Excitation rate coefficients $\langle\sigma\varv\rangle$~for 
different models of inelastic collisions 
with neutral hydrogen atoms and with free electrons,
and their ratios $S$,
as functions of the transition energy and at a fixed temperature.
LCAO refers to \citet{2016PhRvA..93d2705B};
Impulse refers to \citet[][Eq.~9]{1991JPhB...24L.127K};
Impulse-S refers to \citet[][Eq.~18]{1991JPhB...24L.127K};
R-matrix refers to \citet{2007A&amp;A...462..781B};
BSR refers to \citet{2016PhRvA..94d2707T}~(extended
here to higher energies; see \sect{methodcollce}).
The important
$\mathrm{3s\,{^5}S^{o}}\rightarrow\mathrm{3s\,{^3}S^{o}}$~and 
$\mathrm{3s\,{^5}S^{o}}\rightarrow\mathrm{3p\,{^5}P}$~transitions
have been highlighted.}
\label{fig:collisions}
\end{center}
\end{figure*}

\subsubsection{Overview}
\label{methodcolloverview}

The inelastic collisional processes 
in the model atom include 
oxygen plus neutral hydrogen (O+H) excitation,
oxygen plus electron (O+e) excitation,
oxygen plus proton (O+p) charge transfer
(which is efficient 
for the transition between the ground states of \ion{O}{I} and \ion{O}{II}), 
O+e ionisation, and O+H charge transfer,
from most to least influential on the \triplet~strength.
We discuss them in turn below.

Most of the collisional rate coefficient data discussed below
were calculated without resolving fine structure.
To include them into the model atom, which does include fine structure, we note that
it is necessary to divide by the total number of
final states so as to approximately preserve the total rate
per perturber in dimensions of inverse time.
This works because  
the collisional rates within fine structure in our models
were set to very high values, so that 
the population ratios within fine structure levels
are given by the ratios of their statistical weights
\citep[following][]{1993A&amp;A...275..269K}.
This in turn 
can be justified since inelastic collision processes between fine
structure levels are expected to be very efficient on the basis of the
Massey criterion \citep{1949RPPh...12..248M}.

\subsubsection{O+H excitation}
\label{methodcollch}

Inelastic  collisions with neutral hydrogen have by far the most
influence on the emergent \triplet~intensities,
even at disk-centre
(\sect{discussioncollisions}).  Uncertainties in the adopted collisions
therefore dominate the uncertainty in the synthetic spectra,
and consequently in the inferred abundances.
As we discussed in \sect{introduction},
contemporary studies typically use the  Drawin recipe
\citep[][Appendix A, and references therein]{1993PhST...47..186L}.
Here, we do not discuss the Drawin recipe in any detail
other than to point out that this model
contains very little of the relevant
quantum scattering physics \citep[][]{2011A&amp;A...530A..94B}.
We restrict ourselves to more physically motivated recipes
in order to shed light on the true nature of 
inelastic O+H collisions in the solar photosphere.

\begin{itemize}
\item \textbf{LCAO.} 
This refers to the asymptotic two-electron model of
\citet{2016PhRvA..93d2705B}, based on a linear combination of
atomic orbitals (LCAO) approach for the molecular structure (i.e. potentials and
radial couplings) of the O-H quasi-molecule, and the multichannel Landau--Zener
model for the collision dynamics.  It describes electron transfer at avoided
ionic crossings through interaction of ionic and covalent configurations
\citep[see][]{2011A&amp;A...530A..94B,2016A&amp;ARv..24....9B}.  The data were
recently presented in \citet{2017arXiv171201166B}\footnote{Data 
available at \url{https://github.com/barklem/public-data}}, and
preliminary data were employed in \citet{2017A&amp;A...604A..49P}.

\item \textbf{Impulse.} 
This refers to the free electron model of
\citet{1985JPhB...18L.167K,kaulakys1986free,1991JPhB...24L.127K} based on the
impulse approximation.   This model describes the
momentum transfer between the perturbing hydrogen atom and the active
electron (which is assumed to be free) 
on the target (oxygen) atom due to the scattering process between these two
particles. The  \citet{2017ascl.soft01005B}
code was used to evaluate \citet[][Eq.~9]{1991JPhB...24L.127K}, and more details can
be found in \citet{2015A&amp;A...579A..53O}.  The rate coefficients were
redistributed among spin states following \citet[][Eq.~8 and
Eq.~9]{2016A&amp;ARv..24....9B}.

\item \textbf{Impulse-S.}
This also refers
to the free electron model above (Impulse),
but in the scattering length approximation
\citep[][Eq.~18]{1991JPhB...24L.127K}.
This approximation typically agrees with the full model
to within a factor of three, while being approximately
$3\,\dex$~cheaper to compute. 

\end{itemize}

We  show in \sect{discussionmain} that the LCAO model
alone is unable to reproduce the centre-to-limb variation
of the \triplet.
The reason for this can be understood by noting that 
the most important transitions  for  
triplet~formation in the photosphere are
$\mathrm{3s\,{^5}S^{o}}\leftrightarrow\mathrm{3p\,{^5}P}$ and 
$\mathrm{3s\,{^5}S^{o}}
\leftrightarrow\mathrm{3s\,{^3}S^{o}}$~(\sect{discussioncollisions}).  As
pointed out in \citet{2017arXiv171201166B}, 
the relevant avoided crossings for these transitions occur at short range,
and so the LCAO model gives rather low rates for these transitions.
Consequently, the LCAO
model by itself cannot reasonably be expected to give accurate
rates for these transitions because
there are likely to be contributions from mechanisms other than
the radial couplings at avoided ionic crossings.  
In the standard adiabatic
(Born--Oppenheimer) approach, these mechanisms  correspond to rotational and
spin-orbit coupling, and to additional radial couplings.

Compared to the LCAO model, the Impulse model
employs a completely different approach to the structure and scattering problem
and there is no obvious relationship between the two theories
\citep[e.g.][]{1983rsam.book..393F}.
However, what is clear is
that the avoided crossing mechanism is not included in the Impulse model:
the model does not and cannot include ionic configurations because it 
assumes that the active electron on the target (oxygen) atomic nucleus 
is free. 
Thus, the Impulse and LCAO models
do not have any overlap in the described physical mechanism.

Thus, in the absence of detailed full-quantum calculations in the
standard adiabatic approach that include these other mechanisms, one possible
approach is to add the rate coefficients from the LCAO model to those from the
Impulse model (\textbf{LCAO+Impulse}) or,    in order to reduce the computational cost of
calculating the cross-sections, to those from the Impulse-S model (\textbf{LCAO+Impulse-S}).   It should be noted that the Impulse approach contains a number of
approximations which are generally valid only for Rydberg states
\citep[see][]{1983rsam.book..393F}. The application to $n=3$
states  in oxygen is therefore questionable, but we use it to obtain an estimate
of the possible contribution of other mechanisms in the absence of better
alternatives.

We compare the relative magnitudes of the 
different models in \fig{fig:collisions}. 
As expected, the Impulse model predicts much larger rate coefficients
than the LCAO model for the important 
$\mathrm{3s\,{^5}S^{o}}\leftrightarrow\mathrm{3p\,{^5}P}$~and 
$\mathrm{3s\,{^5}S^{o}}
\leftrightarrow\mathrm{3s\,{^3}S^{o}}$~transitions.
Therefore, uncertainties propagating
forward from the  \citet{1991JPhB...24L.127K} recipe 
dominate the overall uncertainties in the non-LTE modelling
of the \triplet.

\subsubsection{O+e excitation}
\label{methodcollce}

After neutral hydrogen,
inelastic collisions with free electrons  have the most influence
on the \triplet.
Contemporary studies usually adopt the
data presented in \citet{2007A&amp;A...462..781B},
based on standard R-matrix calculations 
\citep[e.g.][]{1971JPhB....4..153B,1976AdAMP..11..143B}.
For the first time, we adopt data
based on B-spline R-matrix (BSR) calculations
\citep{2006CoPhC.174..273Z}.
The calculations for oxygen were presented 
in \citet{2016PhRvA..94d2707T},
and extended for this work
to include transitions up to around $12.65\,\mathrm{eV}$~above the 
ground state (up to and including $\mathrm{3s\,^{1}D^{o}}$). 

We compare the two data sets in \fig{fig:collisions}.
The agreement is good with most transitions agreeing to better 
than a factor of two.
For the important
$\mathrm{3s\,{^5}S^{o}}\leftrightarrow\mathrm{3p\,{^5}P}$~transition,
the new data are a factor of two larger at the relevant temperatures.
However, given the overwhelming importance of the O+H excitation collisions
(\sect{discussioncollisions}), adopting the newer data set
only affects the inferred abundances by less than $0.01\,\dex$,
with our adopted model atom and
LCAO+Impulse description for inelastic O+H collisions
(\sect{methodcollch}).

\subsubsection{O+p charge transfer}
\label{methodcollchplus}

The O+p charge transfer rate coefficients 
for the transition coupling the ground states of \ion{O}{I}
and \ion{O}{II} were taken from
\citet{1999A&amp;AS..140..225S}, based on a combination of
various quantal and semi-classical
theoretical calculations and on experimental data.
As pointed out by e.g.~\citet{2015A&amp;A...583A..57S},
this transition ensures that the two levels 
share the same departure coefficients.
Since the ground state of \ion{O}{I} is
guaranteed to be in LTE (by virtue of the 
high ionisation potential of \ion{O}{I})
the effect of this transition is thus to ensure the \ion{O}{II}
level is also in LTE.

\subsubsection{O+e ionisation}
\label{methodcollci}

O+e ionisation rate coefficients were calculated using
the empirical formula from \citet{1973asqu.book.....A}.
Our tests suggest that the formula is accurate to around a factor of two.
In any case, these rates do not play a major role
in the statistical equilibrium of \ion{O}{I},
as we discuss in \sect{discussioncollisions}.

\subsubsection{O+H charge transfer}
\label{methodcollch0}

O+H charge transfer rate coefficients 
were drawn from the same calculations as 
the LCAO model we described in \sect{methodcollch}:
the data were presented in Barklem+, based on the asymptotic two-electron
model presented in  \citet{2016PhRvA..93d2705B}.
\citet{2017A&amp;A...604A..49P} discussed the importance
of these transitions on the faint \triplet~emission
in the lower chromosphere.
In the photosphere, however, these transitions are apparently
much less important (\sect{discussioncollisions}).

\section{Analysis}
\label{analysis}

\subsection{Observations}
\label{analysisobservations}

The quiet-Sun observations of 
\citet{2009A&amp;A...507..417P},
averaged in space and time and normalised by \citep{2009A&amp;A...508.1403P},
were used to analyse the centre-to-limb variation
of the \triplet.
These data were also used in the study of
\citet{2015A&amp;A...583A..57S}, as we discussed in
\sect{introduction}.
They were acquired using the TRI-Port Polarimetric 
Echelle-Littrow (TRIPPEL) spectrograph
\citep{2011A&amp;A...535A..14K}
on the Swedish 1-m Solar Telescope \citep[SST;][]{2003SPIE.4853..341S} 
in May 2007. 

The observations were of five $\mu=\cos\theta$~locations across
the solar disk: $\mu=0.197\pm0.003$,
$0.424\pm0.024$, $0.608\pm0.020$, $0.793\pm0.012$, and $0.999\pm0.001$;
the uncertainty in $\mu$~originates from 
the finite spatial coverage of the slit.
The instrumental profile is approximately Gaussian
with full width at half maximum 
$\varv_{\mathrm{b}}\approx1.50\,\kms$ \citep{2009A&amp;A...507..417P}.

\subsection{Fitting procedure}
\label{analysisfit}

The following analysis is based on directly fitting the
observed spectra, instead of an approach based on 
measured equivalent widths.  
The model spectra were convolved with the instrumental
profile before comparing them to the observed spectra.
All three components of the \triplet\ 
for a given $\mu$-pointing were fit simultaneously.


There are some weak blending lines,
mainly of CN and C$_{2}$, in the region around the \triplet.
\citet[][Sect.~4.1.4]{2009A&amp;A...508.1403P}~reported
that these blends have a small influence on their 
calibration of $\sh$, and, more relevant to this work,
a negligible affect on the oxygen abundances
inferred by profile fitting.
Consequently, we neglect these blends from our analysis,
which is also the approach taken by both
\citet{2009A&amp;A...508.1403P} and 
\citet{2015A&amp;A...583A..57S}.

The  profiles were fit by unweighted $\chi^{2}$ minimisation,
using the IDL routine \texttt{MPFIT} \citep{2009ASPC..411..251M}.
For the 3D non-LTE analyses and a given $\mu$-pointing, 
the main free parameter is the oxygen abundance.
The other free parameter (for a given $\mu$-pointing)
is a global wavelength shift (affecting all three components
of the \triplet~in the same way).
This was necessary to account for
uncertainties in the absolute wavelength calibration 
\citep{2009A&amp;A...508.1403P}.

For the 3D non-LTE analyses, no extra broadening parameters
(microturbulence $\xi$; macroturbulence $\varv_{\mathrm{mac}}$)
were included.  The broadening effects of
the convective velocity field oscillations and temperature inhomogeneities
are implicit in the 3D non-LTE method 
\citep{2000A&amp;A...359..729A}.
However, as we discussed in \sect{methodatmosphereaverage},
the analyses based on the \mtd~model solar atmosphere
also included macroturbulence as a free parameter,
$\varv_{\mathrm{mac}}\left(\mu\right)$.

\subsection{Fits}
\label{analysisfits}

\begin{figure*}
\begin{center}
\includegraphics[scale=0.62]{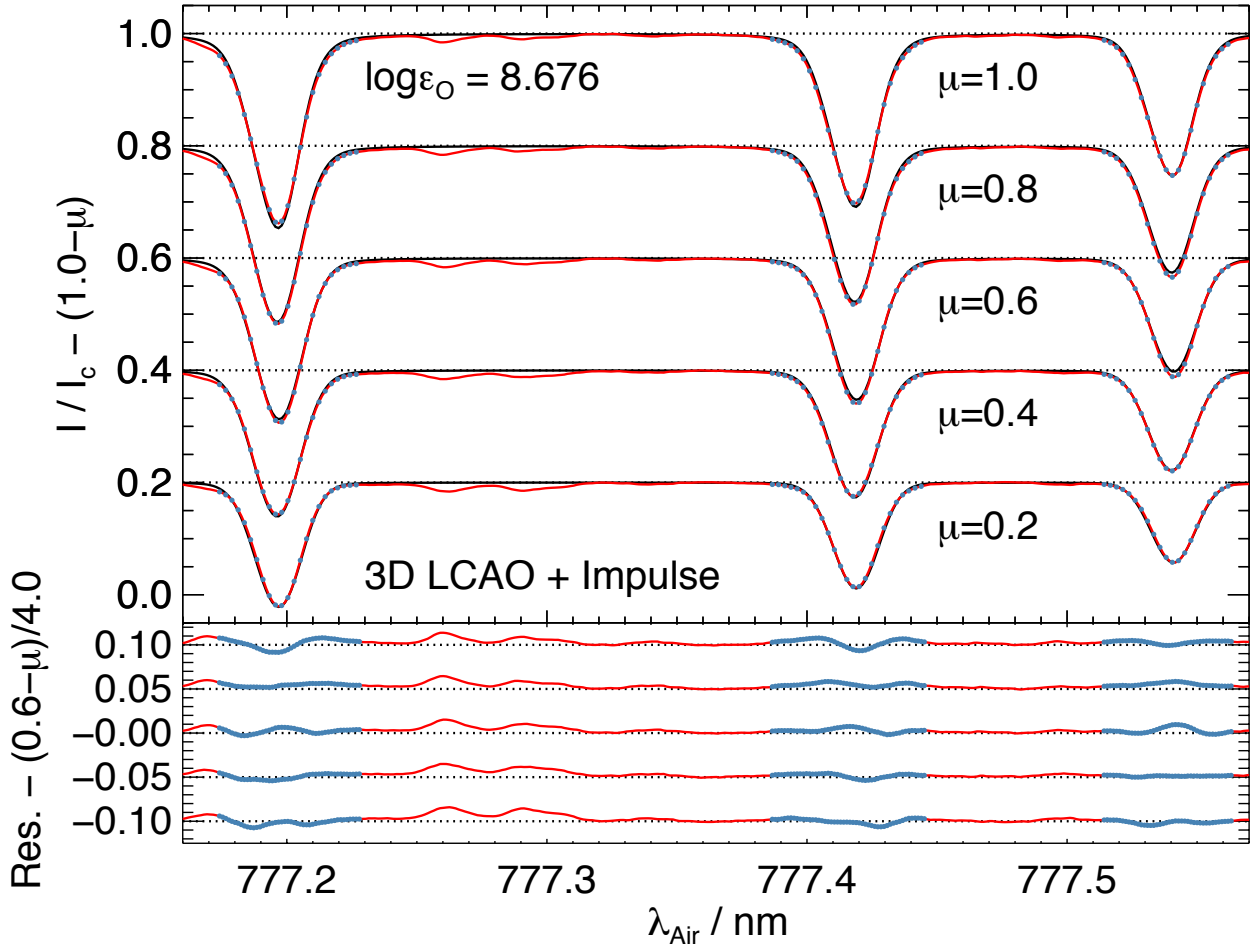}
\caption{Observed (red line) and theoretical (black line) 
centre-to-limb variation of the \triplet~using 
the 3D hydrodynamic model solar atmosphere
and our best description for the inelastic O+H collisions
(\sect{methodcollch}).
The oxygen abundance was fit for $\mu=1.0$,
and fixed to this value for the other $\mu$-pointings.
In the top panel the observational data used in the fit
(blue dots) have been subsampled
by a factor of two to make it easier to see the theoretical spectra
below it. In the bottom panel, the residual is defined as the theoretical
minus the observed spectra.}
\label{fig:clv}
\end{center}
\end{figure*}

\begin{figure*}
\begin{center}
\includegraphics[scale=0.62]{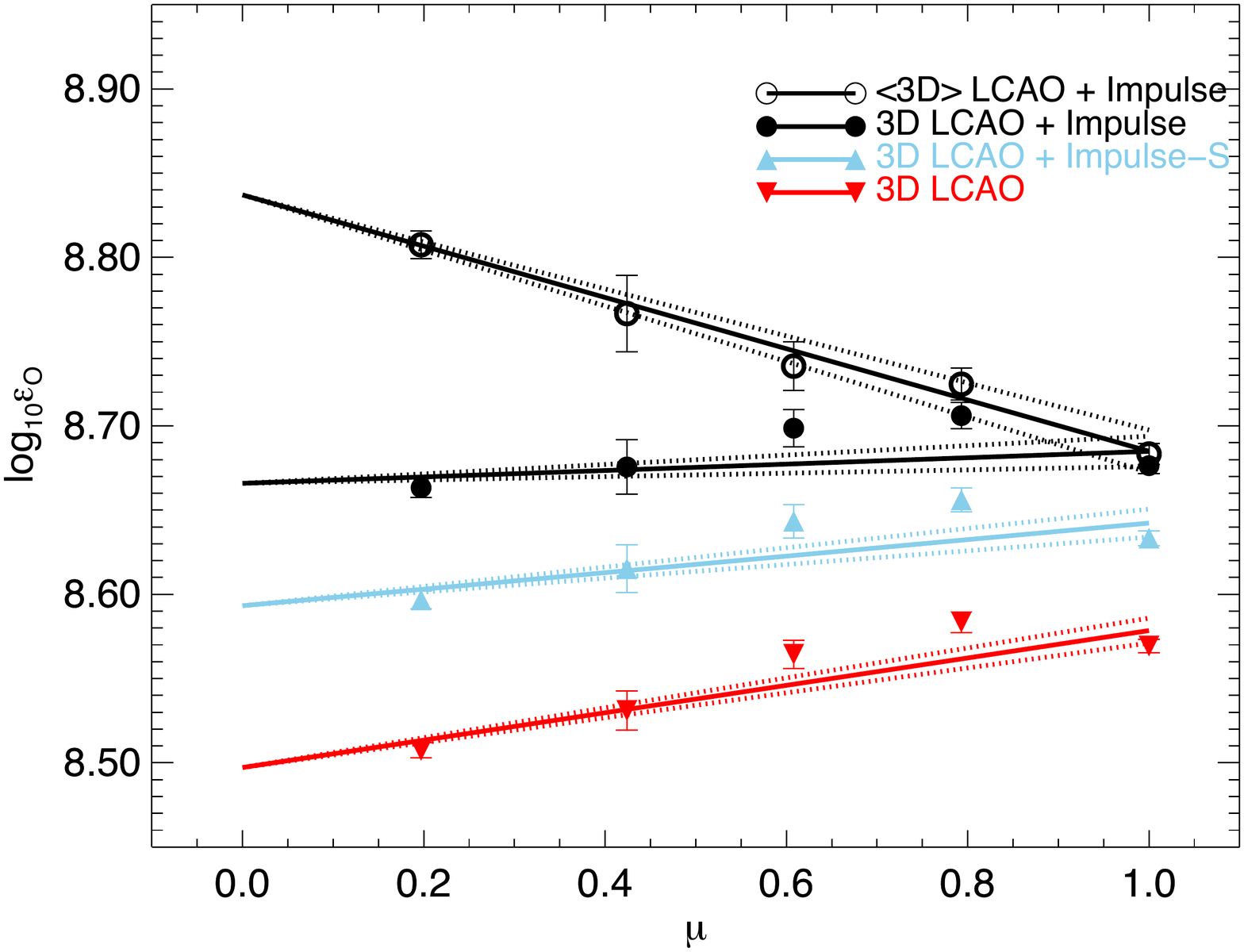}
\caption{Abundances inferred by fitting the observed spectra
at the different $\mu$-pointings individually
using the 3D hydrodynamic model solar atmosphere
and different descriptions for the inelastic O+H collisions
(\sect{methodcollch});
our best description is the LCAO+Impulse model,
based on physical arguments.
The same abundance should be inferred
from each $\mu$-pointing (i.e.~a flat line is expected).
The error bars were calculated by combining in quadrature
the uncertainties arising from
the finite slit-width (or spread in $\mu$),
a $0.1\%$~uncertainty in the continuum placement,
and the formal fitting error;
the first error dominates for
the three intermediate pointings,
while the last two errors are important 
for $\mu\approx0.2$~and $\mu\approx1.0$.
Solid lines of weighted best fit are overdrawn,
and the dashed lines illustrate the $1\,\sigma$~uncertainty
in the fitted gradients.}
\label{fig:fitabund}
\end{center}
\end{figure*}

In \fig{fig:clv}, we illustrate how the observed spectra
compare to the 3D non-LTE model spectra
with the LCAO+Impulse  description
for the inelastic O+H collisions that
we described in \sect{methodcollch}.
The oxygen abundance was fit to the disk-centre spectra,
then fixed to this value for the other $\mu$-pointings.

In \fig{fig:fitabund}, we show the inferred oxygen abundances
after fitting the different $\mu$-pointings separately,
as a function of $\mu$, using the \mtd~model solar atmosphere
with the main, LCAO+Impulse description 
for the inelastic O+H collisions,
and the full 3D model solar atmosphere
with different descriptions
for the inelastic O+H collisions.


\section{Discussion}
\label{discussion}

\subsection{Models versus observations}
\label{discussionmain}

\fig{fig:clv} illustrates that the
3D non-LTE, LCAO+Impulse model reproduces the strengths
and shapes of the lines reasonably well across the solar disk. 
It is apparent, however, that the synthetic lines
are not broad enough. We speculate that this
can be explained by Zeeman broadening,
which is neglected in our purely hydrodynamic simulations.
This imparts a systematic error on our fitted abundances;
however, this error is expected to be small
\citep[of the order of {$0.01\,\dex$}; e.g.][]{2015ApJ...799..150M,
2016A&amp;A...586A.145S}.

\fig{fig:fitabund} demonstrates that
when using the 3D non-LTE, LCAO+Impulse model
for the inelastic collisions (\sect{methodcollch}),
the inferred oxygen abundances are consistent across the
solar disk to a scatter of $\pm0.02\,\dex$. 
The weighted mean abundance is the same as the unweighted mean abundance,
namely $8.684\,\dex$.
The scatter is larger than the
mean trend (the line of best fit), which
gives $8.685\,\dex$~at disk-centre and $8.666\,\dex$~when extrapolated
to $\mu=0.0$,  a discrepancy of just $0.02\,\dex$.

The centre-to-limb discrepancy of $0.02\,\dex$~using
the 3D model solar atmosphere in \fig{fig:fitabund}
could signal that the the LCAO+Impulse model may slightly underestimate
the inelastic O+H collisions overall.
However, we caution that this analysis 
cannot be used to comment on the relative errors between different
transitions. This is why a physically motivated approach is beneficial.
We  would hope, if the same physics describes all of these transitions,
that the relative error between transitions is negligible. 
The scatter of $\pm0.02\,\dex$~about 
the mean trend of centre-to-limb abundances
is inconsistent with the $1\,\sigma$~observational uncertainties.
We stress that the scatter is unlikely to be associated with the
treatment of inelastic O+H collisions,
and that the scatter does not alter our
conclusions about the inelastic O+H collisions.
We note, however, that residual systematic errors such as these
interfere with attempts to calibrate the inelastic O+H collisions 
and $\sh$.

We also cannot completely rule out errors in other aspects of
the modelling and analysis, as the cause of the residual mean trend
and scatter of the order of $0.02\,\dex$. 
For example, the pronounced difference between the 
\mtd~and full 3D trends indicate a sensitivity to the 
model solar atmosphere;
however, the evidence suggests that 
the hydrodynamics of the solar photosphere can be modelled with 
sufficient accuracy, at least in the regions where the \triplet~forms,
as we discussed in \sect{methodatmospherevalidation}.
As we mentioned above, Zeeman broadening is neglected in the models,
and differences of 
about $0.02\,\dex$~may be present.
Finally, these small discrepancies 
could reflect  a systematic error in the observed spectra,
for example in the treatment of stray light or in the accuracy of the 
value of $\mu$~\citep{2009A&amp;A...507..417P}.

We briefly consider the alternative descriptions
for the inelastic O+H collisions.
Using the LCAO+Impulse-S model (\sect{methodcollch}),
the weighted mean abundance drops to $8.627\,\dex$;
 \fig{fig:fitabund} displays
a more prominent trend of decreasing inferred abundance
moving from disk-centre to the limb.
The model spectra are too strong at the limb,
relative to  disk-centre.
Compared to the LCAO+Impulse model above,
this difference can largely be attributed to the 
rate coefficients for the important
$\mathrm{3s\,{^5}S^{o}}\rightarrow\mathrm{3p\,{^5}P}$~transition
(\sect{discussioncollisions})
being a factor of $2.65$~smaller in the Impulse-S
model compared to in the Impulse model (\fig{fig:collisions}).

In \fig{fig:fitabund} 
we also show results calculated using the LCAO model alone. 
The weighted mean abundance drops to $8.547\,\dex$,
which is anomalously low \citep[compared to the abundances inferred from other 
diagnostics; e.g.][]{2009ARA&amp;A..47..481A} and
\fig{fig:fitabund} displays
a very prominent trend of decreasing inferred abundance
moving from disk-centre to the limb.
The model spectra are again too strong at the limb,
relative to disk-centre.
As we discussed in \sect{methodcollch},
this suggests that the mechanism described by the LCAO model,
namely electron transfer at avoided ionic crossings,
is not the dominant mechanism for the important
$\mathrm{3s\,{^3}S^{o}}\rightarrow\mathrm{3s\,{^5}S^{o}}$~and
$\mathrm{3p\,{^5}P}\rightarrow\mathrm{3s\,{^5}S^{o}}$~transitions
(\sect{discussioncollisions}), which occur at short range.

Finally, we briefly comment on the LTE assumption.
For clarity, we do not show the 3D LTE results in \fig{fig:fitabund}.
The inferred abundance at disk-centre in 3D LTE is $8.88\,\dex$,
and this steeply increases towards the limb by around $0.6\,\dex$.
The centre-to-limb variation clearly rules out LTE as 
a valid modelling assumption \citep[e.g.][]{1968SoPh....5..260A}.

\subsection{Sensitivity to the inelastic O+H collisions}
\label{discussioncollisions}

\begin{figure*}
\begin{center}
\includegraphics[scale=0.31]{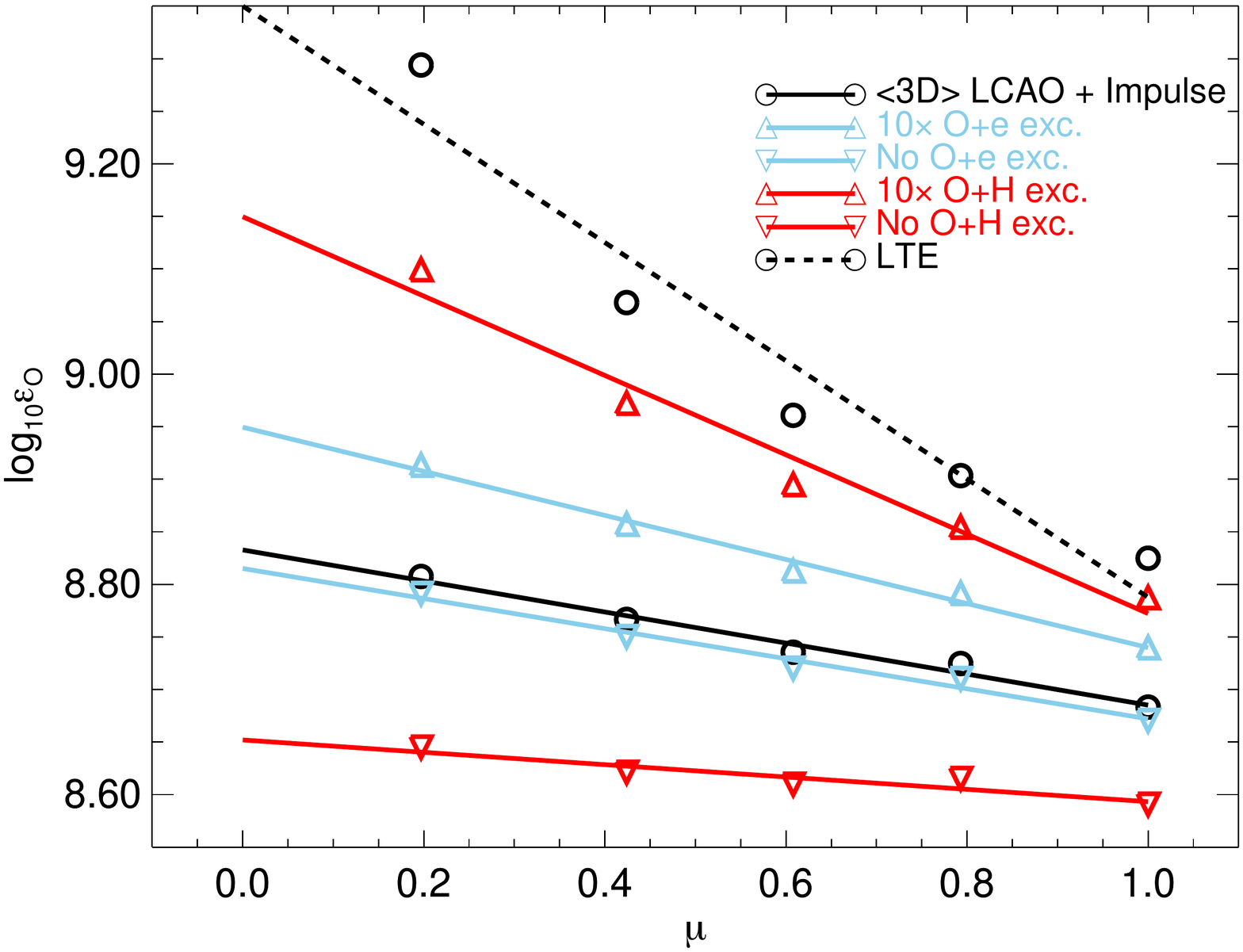}
\includegraphics[scale=0.31]{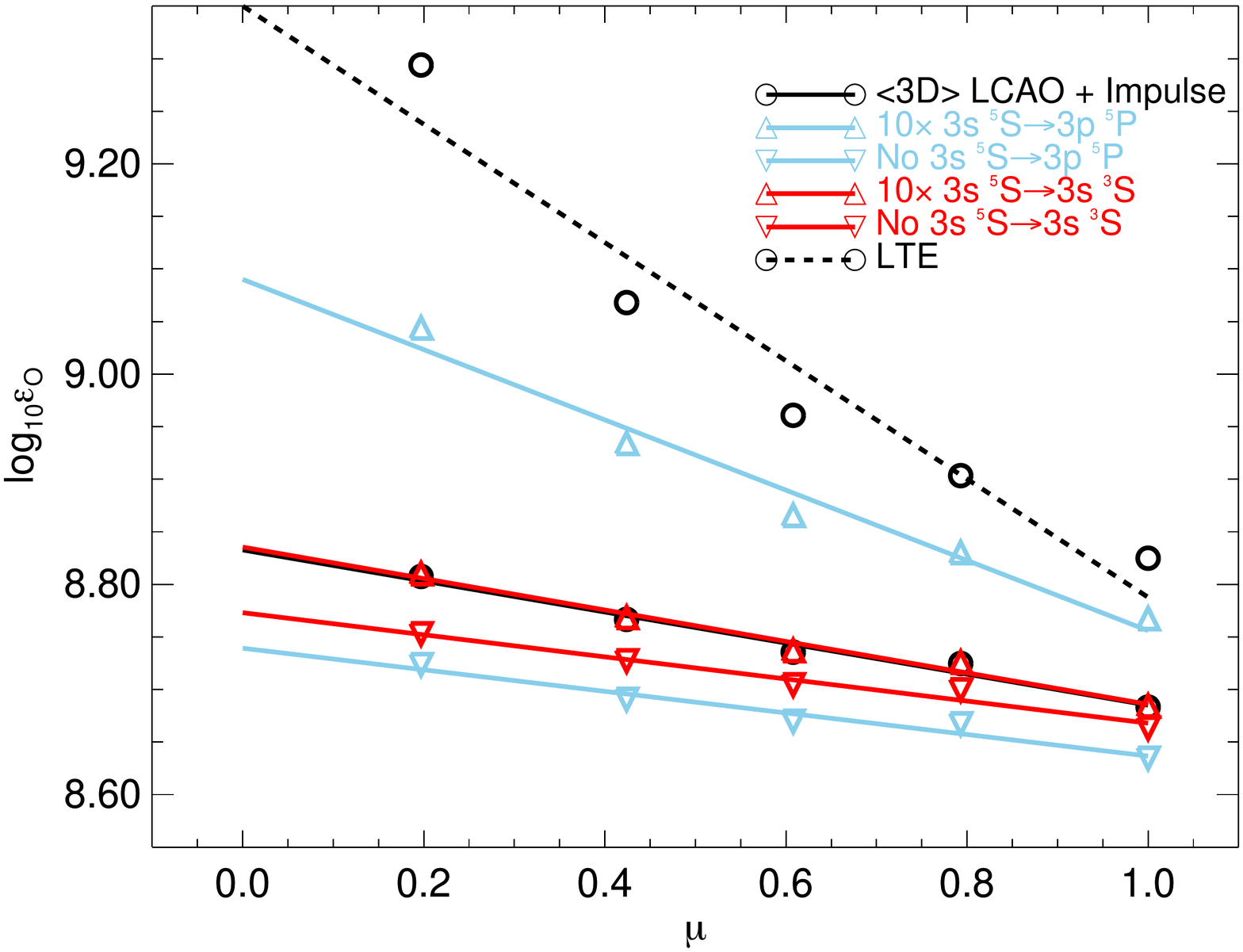}
\caption{Abundances inferred at the different $\mu$-pointings
using the \mtd~model solar atmosphere. 
The same abundance should be inferred
from each $\mu$-pointing (i.e.~a flat line is expected).
The left plot shows the effect of 
altering all inelastic O+e or O+H excitation rate coefficients.
The right plot shows the effect of 
altering the inelastic O+e and O+H excitation rate coefficients
(simultaneously) for the specified transition.}
\label{fig:fitabund_sens}
\end{center}
\end{figure*}

As we explained in \sect{methodatmosphereaverage}, we
used the \mtd~model solar atmosphere to
test the sensitivity of the 
centre-to-limb variation of the \triplet~to 
different aspects of the non-LTE modelling.
We illustrate the sensitivity to different inelastic collisional transitions
in \fig{fig:fitabund_sens}.
The inelastic O+H excitation collisions have the largest
influence on the statistical equilibrium.
Switching them off 
reduces the abundance inferred from the disk-centre spectra
by $0.09\,\dex$,
while enhancing them by a factor of ten
increases the abundance inferred from the disk-centre spectra
by $0.10\,\dex$.

The inelastic O+e collisions have a smaller
but still significant
influence on the statistical equilibrium.
Although switching them off 
reduces the abundance inferred from the disk-centre spectra
by only $0.01\,\dex$, enhancing them by a factor of ten
increases the abundance inferred from the disk-centre spectra
by $0.06\,\dex$.

The other types of inelastic collisional transitions included in the model,
O+p charge transfer, O+e ionisation collisions, 
and O+H charge transfer, have a much smaller impact
on the \triplet.
For clarity, we do not show them in \fig{fig:fitabund_sens}.
Switching them off or enhancing them by a factor of $10$
affects the abundances inferred
from the disk-centre spectra by much less than $0.01\,\dex$.

The most important inelastic collisional transition 
for the \triplet~is the radiatively allowed transition
between $\mathrm{3s\,{^5}S^{o}}$~and $\mathrm{3p\,{^5}P}$~(i.e.
between the levels of the triplet).
The transition directly offsets the photon losses in the 
triplet~\citep{2016MNRAS.455.3735A}.
\fig{fig:fitabund_sens} shows that switching it off reduces
the abundance inferred from the disk-centre spectra 
by $0.05\,\dex$,
while enhancing it by a factor of ten
increases the abundance inferred from the disk-centre spectra
by $0.08\,\dex$.

The second most important inelastic collisional transition 
is the radiatively forbidden, spin-exchange transition between 
$\mathrm{3s\,{^5}S^{o}}$~(i.e.~the lower
level of the \triplet) and $\mathrm{3s\,{^3}S^{o}}$.
It efficiently reduces the overpopulation of the metastable
$\mathrm{3s\,{^5}S^{o}}$~level~\citep[e.g.][]{2017A&amp;A...604A..49P}.
\fig{fig:fitabund_sens} shows that switching it off reduces
the abundance inferred from the disk-centre spectra 
by $0.02\,\dex$.
Enhancing it by a factor of ten
has a negligible impact on the 
triplet~strength; in the present model, the collisions are already efficient enough
to ensure that these two levels share the same departure coefficients.

\subsection{Solar oxygen abundance}
\label{discussionsolarabundance}

We comment on the implications of our results on 
the still disputed solar oxygen abundance.
To infer the solar oxygen abundance from the 
\triplet, it makes sense to use the disk-centre spectra,
since here the observed lines have formed deeper in the atmosphere.
This means that the abundance inferred from the disk-centre spectra
is less influenced by the inelastic O+H collisions
and by departures from LTE, which would otherwise dominate the
overall uncertainty.
Also, the disk-centre spectra and its normalisation has 
already been shown \citep{2009A&amp;A...508.1403P}
to compare well with the high-resolution
disk-centre solar atlas of \citet{1984SoPh...90..205N};
 it is thus unlikely that there
are severe systematic errors in the disk-centre observational data set.

The 3D non-LTE, LCAO+Impulse model gives
a disk-centre abundance of
$8.676\,\dex$~from the \triplet~(\fig{fig:clv}).
However, as we discussed in \sect{discussionmain}, this model gives
a slight gradient in the mean trend of inferred abundances
of about $+0.02\,\dex$~going from the limb to disk-centre.
We make the assumption that the 
inelastic O+H rate coefficients are the dominant source of error;
a reasonable assumption for the Impulse model.
The $\mathrm{3s\,{^5}S^{o}}\rightarrow\mathrm{3p\,{^5}P}$~transition
dominates in importance (\sect{discussioncollisions}).
Separate calculations showed that enhancing the rate coefficient of
the $\mathrm{3s\,{^5}S^{o}}\rightarrow\mathrm{3p\,{^5}P}$~transition
by a factor of two leads to a disk-centre abundance of $8.709\,\dex$,
and the opposite gradient in the mean trend of centre-to-limb abundances
($-0.02\,\dex$).  

Thus, by flattening the residual trend in the 
3D non-LTE, LCAO+Impulse model in this way,
we obtain a recommended
solar oxygen abundance: $\lgeps{O}=8.69\pm0.03$. 
The uncertainty of $0.03\,\dex$~combines the scatter 
about the mean trend of centre-to-limb abundances
from the 3D non-LTE, LCAO+Impulse model ($\pm0.02\,\dex$), 
the uncertainty in this rough calibration of the 
$\mathrm{3s\,{^5}S^{o}}\rightarrow\mathrm{3p\,{^5}P}$~transition
($\pm0.02\,\dex$, half the difference between the two disk-centre
inferred abundances),
errors in the oscillator \triplet~\citep[of 
the order {$\pm0.01\,\dex$};][]{NIST_ASD},
and the errors incurred from neglecting magnetic fields 
\citep[also of the order {$\pm0.01\,\dex$}: e.g.][]{2015ApJ...799..150M,
2016A&amp;A...586A.145S}.

This result agrees with the low solar oxygen abundance
of $\lgeps{O}=8.69\pm0.05$~advocated by \citet{2009ARA&amp;A..47..481A}.
Such a low solar oxygen abundance is still controversial
because it increases the disagreement between 
predictions from standard solar interior models
and helioseismic measurements for the depth of the convection zone, 
the helium abundance in the convective envelope,
and the interior sound speed
in the Sun \citep[e.g.][]{2016EPJA...52...78S,2016JPhCS.665a2078T}.

The cause for this solar modelling problem has been extensively
investigated since the first 3D-based studies implying a low 
($\lgeps{O} \approx 8.70$) solar abundance appeared
\citep{2001ApJ...556L..63A,2004A&amp;A...417..751A}, but 
as  yet without a conclusive resolution. 
Recent findings of substantial 
amount of missing opacity in the solar interior, however, suggest
that this long-standing problem may soon be solved. 
\citet{2015Natur.517...56B}~measured iron opacities 
in conditions close to those near the base of the convection zone
with the Sandia $Z$-pinch machine.  Their measured Rosseland mean opacities 
for iron were a factor of around $1.75$~higher
than predicted 
\citep[by standard opacity models; 
e.g.][]{1996ApJ...464..943I,2003JQSRT..81..227I,2005MNRAS.360..458B}.
This implies that iron alone
can explain about half of the solar modelling problem.
Complementing this, new R-matrix calculations 
found that the Rosseland mean opacity
for the important \ion{Fe}{XVII} ion is enhanced by a factor of around
$1.35$~after including atomic core photoexcitations as well as other 
improvements \citep{2016PhRvL.116w5003N,2016PhRvL.117x9502N,
2018arXiv180102085P}; ensuring completeness of excited configurations 
leads to an additional enhancement of around 
$1.20$~\citep{2018arXiv180102188Z}.
These results suggest that a resolution may be reached without
having to revise the abundance of oxygen (or other key elements)
in the solar atmosphere.


\subsection{Proposed recipe for inelastic X+H collisions}
\label{discussionpropcollisions}

In \sect{methodcollch}, 
we briefly motivated the LCAO+Impulse model
from a theoretical perspective,
and in \sect{discussionmain} we demonstrated that
this model can reproduce the solar centre-to-limb variation
of the \triplet, albeit
with some evidence of a discrepancy
(of about $0.02\,\dex$),
which may indicate that the inelastic O+H collisions are 
slightly underestimated overall.

There is evidence of the validity of the LCAO and 
Impulse models from other studies that have applied them separately.
In particular, \citet{2016PhRvA..93d2705B} showed that the LCAO model
cross-sections for \ion{Li}{I}, \ion{Na}{I}, and \ion{Mg}{I}
compared well with those of detailed full-quantum calculations 
for transitions with high rates. 
The detailed full-quantum calculations in turn agree with experiment 
for the one available case, 
\ion{Na}{I} $3s\rightarrow3p$~\citep{1991JPhB...24.4017F,
1999PhRvA..60.2151B,2011A&amp;A...530A..94B}.
It has previously been demonstrated that the 
latter data sets, when incorporated into non-LTE model atoms,
are needed to accurately model low- and 
intermediate-excitation lines
\citep{2009A&amp;A...503..541L,2011A&amp;A...528A.103L,
2015A&amp;A...579A..53O}.

In contrast, transitions between levels of high excitation
are typically in the regime of the Impulse model.
\citet{2015A&amp;A...579A..53O} 
presented non-LTE \ion{Mg}{I} calculations
based on 1D and \mtd~model atmospheres, 
and was able to reproduce the solar centre-to-limb variation
of high-excitation \ion{Mg}{I} emission lines
after adopting the Impulse model for 
the inelastic Mg+H collisions.

In the absence of detailed 
full-quantum scattering calculations based on 
quantum chemistry calculations of the molecular structure,
it is desirable to have some physically motivated, approximate
description for inelastic X+H collisions to use 
in non-LTE model atoms. The results presented here,
and in the studies discussed above,
signal that the LCAO+Impulse may work to a
satisfactory level of accuracy.
We caution, however, that this needs to be tested
against, for example, the centre-to-limb variation
of other species using 3D non-LTE line formation calculations.
In particular, the Impulse model
employs a number of approximations that increase the
uncertainty of the calculated rate coefficients
\citep[e.g.][]{2016A&amp;ARv..24....9B},
and it may just be coincidental 
that the model performs well for the handful of inelastic O+H transitions
that are most important (\sect{discussioncollisions})
for the line formation of the \triplet.

The computational cost of the 
Impulse model \citep[][Eq.~9]{1991JPhB...24L.127K} can
make it impractical for modelling more complex chemical species such as 
\ion{Fe}{I}/\ion{Fe}{II}. In that case a cheaper alternative may be found
in the Impulse-S model \citep[][Eq.~18]{1991JPhB...24L.127K}. 
The results in \fig{fig:fitabund}~show that
the error is of the order of $0.06\,\dex$~(in terms of oxygen abundance)
for the \triplet.  However,
other lines and species are typically less sensitive to 
the inelastic neutral hydrogen collisions,
so this error of the Impulse-S model relative to the full
Impulse model might be considered an upper bound.
The difference between the Impulse-S model and the Impulse model
may in general be less than the overall uncertainty in
the Impulse model.

\section{Conclusion}
\label{conclusion}

We have presented 3D non-LTE line formation calculations
for the \triplet~on a 3D hydrodynamic \stagger~model solar atmosphere.
For the inelastic O+H collisions, we 
used the asymptotic two electron
model, based on linear combinations of atomic orbitals,
and the free electron model, based on the impulse approximation.
This is more physically motivated than the often-used Drawin recipe,
and may therefore lead to more trustworthy
non-LTE model spectra and abundances.

The 3D non-LTE, LCAO+Impulse model compares
well against the observed solar centre-to-limb variation
of the \triplet.
The mean trend of centre-to-limb abundances
is almost flat, with a gradient
of about $+0.02\,\dex$~going from the limb to disk-centre.
This was achieved without
any calibration of the inelastic collisional rate coefficients,
and without ad hoc line broadening parameters.
Given the success of the LCAO+Impulse model,
and in the absence of detailed quantum chemistry calculations,
we tentatively suggest adopting this approach
for inelastic collisions in non-LTE models of other chemical species.
We caution, however, that this hypothesis needs
to be tested first, for example
by performing analogous centre-to-limb variation
analyses on other chemical species sensitive to 
inelastic collisions with neutral hydrogen.

After flattening the residual trend in the 
3D non-LTE, LCAO+Impulse model,
our recommended solar oxygen abundance from the 
\triplet~is $\lgeps{O}=8.69\pm0.03$,
the error being dominated by systematics.
This strengthens the case for a low solar oxygen abundance.

By combining 3D non-LTE modelling with advances in atomic physics,
we can attempt to model \ion{O}{I} spectral line formation in cool stars 
from first principles.
In the absence of freely varying fudge parameters ($\xi$, $\varv_{\text{mac}}$,
mixing length parameters, van der Waals damping enhancements, $\sh$),
deficiencies inherent in the models become apparent.
In this work, for example, the LCAO model alone 
failed to predict the centre-to-limb variation of the 
\triplet, and that led us to develop the 
improved model presented in this work.
Future work should apply the 3D non-LTE method
to other spectral lines and chemical species to better understand
line-formation from first principles and to 
provide fresh insight into the physics of
planets, stars, and our Galaxy.
As stated by \citet{2017arXiv171201166B}, modern quantum chemistry
calculations including potentials and couplings and detailed scattering
calculations for the low-lying excited states of OH would be very important
in this context.


\begin{acknowledgements}
We thank the anonymous referee for the comments and Hans-G\"unter Ludwig for the discussions
that helped us to improve this manuscript.
AMA acknowledges funds from the Alexander von Humboldt Foundation in the
framework of the Sofja Kovalevskaja Award endowed by the Federal Ministry of
Education and Research.
AMA and MA are supported by the Australian
Research Council (grants FL110100012 and DP150100250).
PSB acknowledges support from the Swedish Research Council and the project grant ``The New Milky Way'' from the Knut and Alice Wallenberg Foundation.
Funding for the Stellar Astrophysics Centre is provided by 
The Danish National Research Foundation (grant DNRF106).
This work was supported by computational resources provided by the Australian
Government through the National Computational Infrastructure (NCI)
under the National Computational Merit Allocation Scheme.
\end{acknowledgements}


\bibliographystyle{aa} 
\bibliography{./bibl}


\label{lastpage}
\end{document}